\documentclass[notitlepage,aps,prd,twocolumn,superscriptaddress,showpacs,nofootinbib]{revtex4-2}

\pdfoutput=1

\usepackage{enumitem}
\usepackage{braket}
\usepackage{hyperref}
\usepackage[bottom]{footmisc}
\usepackage{amsmath,amssymb,amsfonts,mathrsfs}
\usepackage{booktabs}
\usepackage{mathtools}
\usepackage{bm}
\usepackage{dcolumn}
\usepackage{graphicx}   
\usepackage[latin1]{inputenc}
\usepackage{latexsym}
\usepackage{color}
\usepackage[acronym]{glossaries}

\usepackage[caption=false]{subfig}
\newcommand{\hfull}{h^\text{eff}}
\newcommand{\hecho}{h^\T{echo}}
\newcommand{\hnorm}{h^\infty}

\newcommand{\Hp}{\mathscr H^+}
\newcommand{\Teco}{\mathcal{T}^{\text{ECO}}}
\newcommand{\Reco}{\mathcal{R}^{\text{ECO}}}
\newcommand{\RBH}{\mathcal{R}^{\text{BH}}}
\newcommand{\RQBH}{\mathcal{R}^\T{QBH}}
\newcommand{\Hm}{\mathscr H^-}
\newcommand{\Zecho}{Z^{\text{echo}}_{\ell m \omega}}

\newcommand{\Zin}{Z^{\text{in}}_{\ell m \omega}}
\newcommand{\Zouteco}{Z^{\text{out ECO}}_{\ell m \omega}}

\newcommand{\Zoutqbh}{Z^{\text{out QBH}}_{\ell m \omega}}

\newcommand{\ZBH}{Y^{\text{horizon}}_{\ell m \omega}}
\newcommand{\Zout}{Z^{\text{out}}_{\ell m \omega}}
\newcommand{\Din}{D^{\text{in}}_{\ell m \omega}}
\newcommand{\Dout}{D^{\text{out}}_{\ell m \omega}}
\newcommand{\Cin}{C^{\text{in}}_{\ell m \omega}}
\newcommand{\Cout}{C^{\text{out}}_{\ell m \omega}}
\newcommand{\Zinf}{Z^{\infty}_{\ell m \omega}}
\newcommand{\Yin}{Y^{\text{in}}_{\ell m \omega}}
\newcommand{\Yineco}{Y^{\text{in ECO}}_{\ell m \omega}}
\newcommand{\Yinqbh}{Y^{\text{in QBH}}_{\ell m \omega}}
\newcommand{\Yout}{Y^{\text{out}}_{\ell m \omega}}
\newcommand{\Yinf}{Y^{\infty}_{\ell m \omega}}
\newcommand{\T}[1]{\text{#1}}

\newcommand{\scrip}{{\mathscr I}^+}
\newcommand{\scrim}{{\mathscr I}^-}
\newcommand{\RNum}[1]{\uppercase\expandafter{\romannumeral #1\relax}}

\newcommand{\dd}{\mathrm{d}}
\newcommand{\hmem}{h_\text{mem}}

\makeglossaries
    
\newacronym{e2e}{E2E}{End-To-End}
\newacronym{inrep}{INREP}{Initial Noise REduction Pipeline}
\newacronym{tdi}{TDI}{Time Delay Interferometry}
\newacronym{ttl}{TTL}{Tilt-To-Length couplings}
\newacronym{dfacs}{DFACS}{Drag-Free and Attitude Control System}
\newacronym{ldc}{LDC}{LISA Data Challenge}
\newacronym{lisa}{LISA}{Laser Interferometer Space Antenna}
\newacronym{emri}{EMRI}{Extreme Mass Ratio Inspiral}
\newacronym{ifo}{IFO}{Interferometry System}
\newacronym{grs}{GRS}{Gravitational Reference Sensor}
\newacronym{tmdws}{TM-DWS}{Test-Mass Differential Wavefront Sensing}
\newacronym{ldws}{LDWS}{Long-arm Differential Wavefront Sensing}
\newacronym[	plural={MOSAs},
		        first={Moving Optical Sub-Assembly},
		        firstplural={Moving Optical Sub-Assemblies}
            ]{mosa}{MOSA}{Moving Optical Sub-Assembly}
\newacronym{siso}{SISO}{Single-Input Single-Output}
\newacronym{mimo}{MIMO}{Multiple-Input Multiple-Output}
\newacronym[plural=MBHB's, firstplural=Massive Black Holes Binaries (MBHB's)]{mbhb}{MBHB}{Massive Black Holes Binary}
\newacronym{cmb}{CMB}{Cosmic Microwave Background}
\newacronym{sgwb}{SGWB}{Stochastic Gravitational Waves Background}
\newacronym{pta}{PTA}{Pulsar Timing Arrays}
\newacronym{gw}{GW}{Gravitational Wave}
\newacronym{snr}{SNR}{Signal-to-Noise Ratio}
\newacronym{gr}{GR}{General Relativity}
\newacronym{bbh}{BBH}{Binary Black Hole}
\newacronym{bh}{BH}{Black Hole}
\newacronym{pbh}{PBH}{Primordial Black Holes}
\newacronym{eco}{ECO}{Exotic Compact Objects}
\newacronym{qbh}{QBH}{Quantum Black Holes}
\newacronym{qnm}{QNM}{Quasi-Normal Modes}

\newcommand{\CornellPhysics}{\affiliation{Department of Physics, Cornell University, Ithaca, NY, 14853, USA}}
\newcommand{\Cornell}{\affiliation{Cornell Center for Astrophysics and Planetary Science, Cornell University, Ithaca, New York 14853, USA}}
\newcommand{\CornellLepp}{\affiliation{Laboratory for Elementary Particle Physics, Cornell University, Ithaca, New York 14853, USA}}
\newcommand{\Caltech}{\affiliation{Theoretical Astrophysics 350-17, California Institute of Technology, Pasadena, CA 91125, USA}}

\newcommand{\orcid}[1]{\href{https://orcid.org/#1}{\includegraphics[width=8pt]{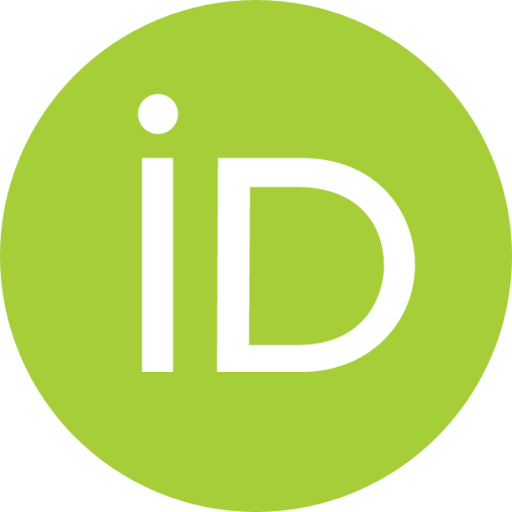}}}

\begin{document}

\title{Signatures of Quantum Gravity in Gravitational Wave Memory}

\author{Nils Deppe \orcid{0000-0003-4557-4115}} \CornellLepp \CornellPhysics \Cornell

\author{Lavinia Heisenberg}
\affiliation{Institute for Theoretical Physics, University of Heidelberg, 
D-69120	Heidelberg,
Germany} 

\author{Lawrence E.~Kidder \orcid{0000-0001-5392-7342}} \Cornell

\author{David Maibach \orcid{0000-0002-5294-464X}}
\email{d.maibach@thphys.uni-heidelberg.de}
\affiliation{Institute for Theoretical Physics, University of Heidelberg, 
D-69120	Heidelberg,
Germany} 

\author{Sizheng Ma \orcid{0000-0002-4645-453X}}
\affiliation{Perimeter Institute for Theoretical Physics, Waterloo, ON N2L2Y5, Canada} 

\author{Jordan Moxon \orcid{0000-0001-9891-8677}} \Caltech
\author{Kyle C.~Nelli \orcid{0000-0003-2426-8768}} \Caltech
\author{William Throwe \orcid{0000-0001-5059-4378}} \Cornell
\author{Nils L.~Vu \orcid{0000-0002-5767-3949}} \Caltech

\date{\today}


\begin{abstract}
We study the impact of quantum corrections to gravitational waveforms on the gravitational wave memory effect. In certain quantum gravity theories and semi-classical frameworks, black holes (or other exotic compact objects) exhibit reflective properties that cause quasi-normal modes of a binary merger waveform to partially reflect off the horizon. If these reflections reach the detector, the measured gravitational wave signal may show echo-like features following the initial ringdown phase. Detecting such echoes, or their indirect signatures, would offer compelling evidence for the quantum nature of black holes. Given that direct detection of echoes requires finely tuned waveform templates, exploring alternative imprints of this phenomenon is crucial.
In this work, we pursue this goal by calculating corrections to the null memory arising from echo-like features, formulated in terms of the Newman-Penrose scalar $\Psi_0$. We demonstrate that the morphology of the resulting features is model-independent rendering them conceptually much easier to detect in real interferometer data than the raw echo. The corresponding signal-to-noise ratio of echo-induced features appearing in the gravitational wave memory is estimated subsequently. We further compute the physical fluxes associated to the echo at both the black hole horizon and null infinity and identify novel distinguishing features of the underlying reflectivity models in measurement data.

\end{abstract}



\maketitle


\newcommand{\myhyperref}[1]{\hyperref[#1]{\ref{#1}}}


\section{Introduction} 
\label{sec:intro}
The (non-linear) \gls{gw} memory effect \cite{Lin_Mem_II, Lin_Mem_I, Lin_Mem, Zeldovich_mem, Christodoulou_mem, Throne_mem, Favata_mem, Blanchet_mem, Bieri_mem}, representing a physical manifestation of the nonlinearity intrinsic to \gls{gr}, is a particularly interesting aspect of the theory and has yet to be confirmed experimentally. This memory reveals itself by a permanent displacement of the freely floating test masses within an idealized \gls{gw} detector after it has been traversed by gravitational radiation. It has been subject to many investigations and plays a key role, for instance, in the Bondi-Metzner-Sachs formalism of asymptotically flat spacetimes \cite{Bondi_Origin_2,Sachs_Origin_2,Sachs_Origin_3}, Weinbergs's soft theorems \cite{Weinberg_ST}, and the infrared triangle \cite{Strominger:2016, strominger_2018_lectures}. The memory carries a manifold of analytically and observationally interesting phenomenology, ranging from connections to the fundamental symmetries of spacetime (as established by Bondi and others) to intrinsic (kinematic) properties of \gls{bbh} mergers \cite{Favata_2009}. Beyond that, however, it can also pick up more subtle features such as corrections due to modified gravity theories \cite{Heisenberg_2023, Brans_Dicke_I, Brans_Dicke_II, Brans_Dicke_III, Chern_Simons_I, Chern_Simons_II, Ma:2024bed}. One aspect that has only sparsely been addressed in literature is the remnant of potential \gls{bh} quantum features in the \gls{gw} memory. While there is a small body of works aiming to treat gravitational radiation from a quantum mechanical point of view, for instance \cite{Parikh_2021, Guerreiro_2022}, the quantumness in \gls{gw} physics comes up majorly in the context of the \gls{gw} echo effect \cite{Cardoso_2016,Cardoso_2016:I,Cardoso_2016:II,Cardoso_2017,Cardoso_2019,Agullo_2021} (see also \cite{Afshordi_I,Afshordi_II,Damico_2020,Manikandan_2022, Chakraborty_2022}). As will be demonstrated in this work, the echo induces a plentiful of features within gravitational waveforms, including a contribution to the memory. Generally, the echo effect refers to residual \gls{gw} signals from a \gls{bbh} merger arriving at the detector with a constant (but potentially frequency-dependent) time delay after the primary waveform has passed through. This echo-like signature is produced when the ingoing radiation from the ringdown phase of a binary merger is reflected by a (quantum) surface close to the classical event horizon. A delayed GW echo is predicted by a range of theoretical models, including those involving deviations from \gls{gr}\cite{Zhang_2018,Dong_2021}. In this work, the cases of interest are echoes arising due to the presence of near-horizon (quantum) structure surrounding the \gls{bh} \cite{Almheiri_2013, Giddings_2016, Oshita_2019, Abedi_2023,Chakravarti_2021,Cardoso_2019,Wang_2020, Oshita_2020} as well as the existence of \gls{eco} replacing the traditional \gls{bh} by substituting the would-be horizon with a physical boundary \cite{Mazur_2004,Visser_2004,Damour_2007,Mathur_2005, Holdom_2017, Mark_2017}. In particular in the context of near-horizon quantum structure, the echo manifests inadmissible evidence of an interplay between quantum physics and gravity \cite{Chakraborty_2022, Oshita_2020, Oshita_2019}. Recent studies also suggest the possibility of echoes arising due to simple quantization arguments, containing imprints of the specific underlying theory of quantum gravity \cite{Agullo_2021}. \gls{eco}s on the other hand challenge the very foundations of our understanding of compact stellar objects and provide new avenues to resolve long-standing conceptual issues that have been present in \gls{bh} physics.
Despite the existence of stable \gls{eco}s and their production of echoes being under debate (see \cite{Ma_2022} and references therein), searching for \gls{gw} echoes in upcoming measurement data has the potential to enlighten the fundamental dark spots in our current understanding of \gls{eco}s and, in particular, (quantum) BHs.

The variety of phenomenological models sourcing a \gls{gw} echo generally complicates the computation of contribution in an agnostic manner. Nonetheless, the prospect of new data collected by future space- and ground-based instruments has facilitated the efforts towards constructing accurate gravitational waveforms including echoes \cite{Surrogate_2021,Mukherjee_2022,Ma_2022}. Attaching an echo to the existing waveform model templates further increases the already large parameter space for numerical waveform simulations. Even for echo-less waveforms, more efficient waveform modeling is already an active research topic as the demands for the precision of such models will increase for future \gls{gw} data \cite{Ashtekar_2019, DAmbrosio_2024}. Adding another layer of complexity to the template construction process by attaching a model-dependent echo to the waveform seems counterproductive. One can therefore ask, whether simulating the exact shape of the echo time series can be avoided without losing important astrophysical and \gls{bh} related information, or, in other words, does the echo also leave traces in the waveform which are less (or not at all) sensitive towards the particular assumptions that determine its the echo's morphology? 

A common feature of all strain signals passing through a \gls{gw} detector is that they carry gravitational energy. This implies that for each signal, there exists a finite interval in the strain time series $h(t)$ during which $|\dot{h}(t)|^2>0$. Even though the precise form of the strain time series may be uncertain due to precision limitations in the data analysis or regarding the waveform templates, the features appearing in detector data due to the passing energy flux are well-defined: The flux through the detector will cause a contribution to the non-linear memory, leading to a permanent displacement of the (ideals) detector's test masses. Consequently, each echo contributes to the accumulated \gls{gw} memory, simplifying the complex time series into a step-like increase in the measured strain. In this work, we explore this claim and estimate the detectability of echo-induced memory, examine the information that can be extracted from such observations, and investigate how these findings are connected to the dynamics that generate \gls{gw} echoes.

To date, neither \gls{gw} memory nor the echo have been detected. Regarding the memory, the measurements of current ground-based have not achieved sufficient \gls{snr} for a claimed detection, as they typically respond to \gls{gw}s with timescales much shorter than those of typical memory signals. However, future ground-based detectors \cite{future_ground_based_I, future_ground_based_II, future_ground_based_III, future_ground_based_IV, future_ground_based_V} and space-based missions \cite{space_based_I,space_based_II} hold great potential for detecting the \gls{gw} memory. In particular, the \gls{lisa} \cite{LISA}, with its freely-floating test masses and excellent sensitivity in the low-frequency band, where memory signals tend to be stronger, stands out as a promising candidate to probe the memory effect. Indeed, early estimates indicate a promising \gls{snr} for massive binary systems at redshifts up to $z\approx 2$ \cite{Henri_mem}. Similarly, while there has been no direct detection of \gls{gw} echoes despite numerous efforts (see \cite{abedi_2020,abedi2020echoesabyssstatusupdate} for a review and also \cite{Abedi_2019}), \gls{lisa}'s prospects for detecting echoes have been demonstrated to be promising in a companion paper \cite{Maibach_2024}. These prospects remain strong even when echoes are treated as separate signals, detached from their ``sourcing'' waveform - a scenario that has been suggested as plausible \cite{zimmerman2023rogueechoesexoticcompact}. 

Building up on the tentative findings for echoes in \cite{Maibach_2024}, we begin our investigation into the echo-induced memory by examining the reflective properties of \gls{eco}s and \gls{bh}s in Section \ref{sec:theoretical_background}. For the \gls{bh} reflectivity, we rely on the model developed in \cite{Maibach_2024}, which is based on area quantization arguments and referred to as \gls{qbh} throughout this work. For \gls{eco}s, we use the model adopted in \cite{Ma_2022}. Section \ref{subsec:echo_reconstruction} details the methodology for constructing echo templates from numerical waveforms, comparing the results for both reflectivity models. In Section \ref{sec:balance_laws}, we explore the properties of echo-induced memory and derive semi-analytical expressions for its dominant components. Additionally, we investigate the associated flux balances. To complement the analytical findings, we conduct numerical simulations to explore the echo-induced memory and its related flux laws, utilizing a set of waveforms from the \textit{SXS} dataset (see \cite{Boyle_2019} for the latest update). Distinctive features of memory contributions resulting from different theoretical origin are explored. Furthermore, we estimate the \gls{snr} for the memory contributions due to echoes using a recently developed \gls{lisa} data analysis pipeline \cite{Henri_mem}. A comprehensive discussion of the numerical results and their implications is provided in Section \ref{sec:conclusion}.



\section{Gravitational Waves from Quantum Black Holes: Theory}
\label{sec:theoretical_background}

The mechanism leading to the creation of GW echoes is based on a perturbative treatment of the spacetime metric outside the compact object formed during a binary merging event. After the merger, the remnant body adopts a perturbed Kerr (or Schwarzschild) metric, which gradually settles into an unperturbed state. The decay of these perturbations is governed by the Teukolsky equations \cite{Teukolsky_1972,Teukolsky_1973}. The latter can be solved analytically, revealing both an ingoing and outgoing solution in terms of \gls{qnm}s. The outgoing solution corresponds to the ringdown phase of the gravitational waveform observed by the GW detector. 
The ingoing solution, in the classical framework, crosses the event horizon of the BH and is lost forever inside the horizon. This part of the solution is typically considered irrelevant in classical treatments of BH physics. However, this picture changes dramatically if the horizon is no longer treated as a fully transmissive, ``unspectacular'' surface.

\subsection{Reflective Properties of Quantum Black Holes and Exotic Compact Objects}
\label{subsec:quantum_bh}
In the quantum regime, the BH horizon plays a significant role for physical processes, as it has first been established by Hawking when he demonstrated that the BH in fact emits thermal radiation. 
To date, the exact quantum structure of a BH's horizon remains unknown. Nevertheless, numerous theories have been proposed to model its quantum properties, including the Membrane Paradigm \cite{Damour_book, Throne_book, PhysRevD.33.915} (see also \cite{Maggio_2020, Chakraborty_2022,Rahman_2021}), effective field theories \cite{Burgess_2018}, phenomenological approaches \cite{Oshita_2020} and quantum mechanics-inspired model \cite{PhysRevD.105.044046}.
In its original form, the Membrane Paradigm postulates a replacement of the BH horizon by a fictitious (physical) membrane. This framework has been particularly useful for modeling the outer layers of ECOs. In this article, we focus on two recent proposals in which i) the Membrane Paradigm is used to impose boundary conditions near an ECO's surface \cite{Oshita_2020, Wang_2020, Chakraborty_2022} and a Boltzmann reflectivity applies (see also \cite{Mark_2017} for similar approaches avoiding Membrane Paradigm), and ii) a BH reflectivity is constructed based purely on area quantization arguments \cite{Agullo_2021, Maibach_2024}. We outline the relevant results from both approaches in the following sections.

\subsubsection{Boltzmann Reflectivity and Membrane Paradigm}
\label{subsub:Boltzman}
Replacing the BH horizon with a (quantum) membrane fundamentally alters the behavior of linear perturbations of the BH spacetime. In \cite{Oshita_2020} it was argued that quantum effects near the horizon, which lead to thermal emission at the Hawking temperature $T_\T{H}$, modify the equations governing the evolution of these linear perturbations. Concretely, the Membrane Paradigm and other effects in convolution with the fluctuation-dissipation theorem lead to
\begin{align}
\label{equ:fluctuation_dissipation}
    \left(-i\frac{\gamma \Omega(x)}{E_\T{Pl}}\frac{\dd^2}{\dd x^2} + \frac{\dd^2}{\dd x^2} + \omega^2 -V(x)\right)\psi_\omega(x)=\xi_\omega(x)\,,
\end{align}
as the evolution equation for QNM functions, $\psi_\omega$, of GWs under a stochastic fluctuation field $\xi_\omega$. In the latter, $\Omega(x):= \omega/\sqrt{g_{00}}$ is the blueshifted frequency, $E_\T{Pl}$ the Planck energy, $\gamma$ the dimensionless dissipation parameter, and $V(x)$ the \gls{bh} potential. The dissipation term resembles the viscous dissipation for sound waves but is suppressed by the gravitational coupling $\Omega/E_{Pl}$. 
The latter nicely matches with phenomenology resulting from quantum gravity theories where constraints on spacetime viscosity can be implemented via viscous terms \cite{Liberati_2014}. 

Following \eqref{equ:fluctuation_dissipation}, the fluctuation-dissipation theorem predicts a thermal spectrum for the mode functions $\psi_\omega$. Near the horizon, one can thus compute the analytical solution of the QNM functions assuming a constant surface gravity of $\kappa = 2\pi T_\T{H}$. Physically, this can be interpreted as the energy flux of ingoing GWs being unable to penetrate the BH horizon and getting either absorbed or reflected. This result is consistent with the models for which the BH interior cannot be entered, such as the Membrane Paradigm or the String Theory inspired fuzzball complementary conjecture \cite{Mathur_2014,Chowdhury_2013}. 

In the near horizon limit, $\-\log(E_{Pl}/\gamma|\omega|)\ll \kappa x\ll-1$, one thus finds the Boltzmann (flux) reflectivity \cite{Oshita_2020}
\begin{align}
\label{equ:boltzmann_ref}
    |\mathcal{R}| = \left| \frac{e^{-\pi|\omega|/(2\kappa)}\bar Y}{e^{\pi|\omega|(2\kappa)} Y} \right|^2 = e^{-|\omega|/(2T_\T{H})}\,,
\end{align}
We note this equation is independent of the dissipation parameter $\gamma$ appearing in the dissipation term of equation \eqref{equ:fluctuation_dissipation}. This is due to taking the absolute value. The parameter $\gamma$ determines the time-separation of the reflected echoes \cite{Oshita_2020} and thus enters as a complex phase. Concretely, including dissipation effects, the position of reflection is marked by the distance $x_0$ for which $\gamma\Omega(x)\sim E_\T{Pl}$. Thus, in this model,
\begin{align}
\label{equ:t_echo_I}
    \Delta t_\T{echo} = 2|x_0| = -\frac{\ln{(\gamma|\omega|)}}{\pi T_\T{H}}\,.
\end{align}
For a more exhaustive derivation of the latter, we refer to \cite{Wang_2020}. The time delay can be rewritten as a complex phase and incorporated into $\mathcal{R}$ such that
\begin{align}
\label{equ:reflectivity}
    \mathcal{R} = \exp{\left(-i\omega\frac{\ln{(\gamma|\omega|)}}{\pi T_\T{H}}\right)}\exp{\left(-\frac{|\omega|}{2T_\T{H}}\right)}\,.
\end{align}
Note here that while the arguments leading to the Boltzmann reflectivity are less model-dependent, the phase of reflectivity results purely from equation \eqref{equ:fluctuation_dissipation}, in particular, the structure of the dissipation term. Thus, the time delay between echoes is highly sensitive to the underlying (quantum) theory\footnote{In \cite{Chakraborty_2022}, for instance, the Membrane Paradigm is treated more rigorously and via a replacement of the horizon with a fictitious quantum membrane. This membrane constitutes an ensemble of microscopic degrees of freedom in the ground state following a Gaussian wave function.
As a result, a frequency-dependent reflectivity can be defined (see Fig. 1 in \cite{Chakraborty_2022}) and similarly, a time separation between echoes is defined as $\Delta t = 2M[(1-\sigma/\sqrt{\pi}-2\log{(\sigma/\sqrt{\pi})}]$. Both the reflectivity and $\Delta t$ are largely dependent on the variance of the quantum state describing the microscopic degrees of freedom, $\sigma^2$.}. Similar results to \cite{Oshita_2020,Wang_2020,Chakraborty_2022} have also been obtained following approaches with very different physical motivation, e.g., \cite{Burgess_2018}.
In this work, the reflectivity \eqref{equ:reflectivity} will serve as a toy model for the reflective properties of ECOs. Thus, we henceforth label equation \eqref{equ:reflectivity} with $\Reco$. We generalize the ECO's reflectivity slightly by introducing an effective horizon temperature $T_\T{QH}$, in place of the standard Hawking temperature $T_\T{H}=1/8\pi$. This allows us to explore how varying the horizon temperature influences the properties of GW echoes, providing insights into the impact of different quantum horizon effects on the observed signals. 

Physically, the first factor in equation \eqref{equ:reflectivity} represents the position of the membrane located near the horizon from which the GW reflects. In radial coordinates, this location can be expressed as $r_\T{echo}\sim\frac{\ln{\gamma}}{2\pi T_\T{QH}}$. The second exponential factor in \eqref{equ:reflectivity} governs the frequency range over which the reflectivity is significantly large. Thus, while the dissipation factor $\gamma$ characterizes the time separation of the echoes, the effective horizon temperature $T_\T{QH}$ primarily determines the width of the reflected frequency band. Generally, the exponential suppression of larger frequencies is motivated by a quantum mechanical perspective, treating an isolated BH as a multilevel system \cite{Oshita_2020}. In this ``giant atom'' model, Hawking radiation serves as the mechanism through which the BH can (spontaneously) de-excite, while the BH's reflectivity can be interpreted as a form of stimulated emission.  Consequently, the frequency scale $\omega \lesssim T_\T{H}$ naturally emerges as the regime where stimulated emission is possible, while for frequencies $\omega \gg T_\T{H}$ the BH behaves as effectively \textit{all-absorbing}.


\subsubsection{Area Quantization and Black Hole Energy States}
\label{subsub:area_quant}
Even when ECOs are not permissable\footnote{An detailed discussion on this matter is presented in \cite{abedi_2020}.}, simple quantization arguments can still support the existence of GW echoes.
A compelling proposition was made in \cite{Agullo_2021}, which is based on Bekenstein and Mukhanov's area quantization formula $A_N = \alpha \ell_\T{Pl} N$, where $\ell_\T{Pl}$ is the Planck length and $N$ a positive integer. This approach leads to a discrete mass spectrum for a given BH, which in turn quantizes its emission and absorption processes. Considering the spin of the BH - though we generally focus on non-spinning remnant BH with dimensionless angular momentum $a\ll1$ - and the fact that, in astrophysical context, we encounter macroscopic BHs only, i.e., $N\gg1$, one finds for the characteristic frequencies being absorbed by the BH \cite{Agullo_2021} that
\begin{align}
\label{equ:characteristic_frequency}
    \omega_N = \frac{\kappa \alpha N}{8\pi} + 2\Omega_H + \mathcal{O}(N^{-1})\,.
\end{align}
Here, $\kappa = \sqrt{1-a^2}/2M(1+\sqrt{1-a^2})$ and $\Omega_H = a/2M(1+\sqrt{1-a^2})$. The coefficients $\kappa$ and $\Omega_H$ denote the surface gravity and the angular momentum, respectively. With the frequency $\omega_N$ scaling as $1/M$, Planck-scale effects are magnified and elevated into the frequency regime relevant for present-day GW interferometers. 

In principle, the characteristic frequencies serve as the sole, narrow pathway for GWs to enter a \gls{bh} by crossing its horizon.  However, if the \gls{bh} is spinning rapidly enough, the width of the quantized energy levels, denoted as $\Gamma$, becomes significant. This width is inversely proportional to the decay rate associated with the spontaneous emission of Hawking radiation, which leads to the de-excitation of the \gls{bh} \cite{Agullo_2021}. If the energy states were to overlap, BHs would behave as true absorbers, allowing any frequency $\omega$ to cross the horizon. This scenario, however, does not hold, even for highly spinning remnant BH, as long as the phenomenological constant $\alpha$ exceeds a critical value $\alpha>\alpha_\T{crit}$. Notably, this critical value $\alpha_\T{crit}$ is much smaller than the lowest phenomenological constant typically considered in the literature \cite{Agullo_2021}, i.e., is $\alpha = 4\log 2$. Consequently, the overlap of energy levels is generally not anticipated and even for $\alpha = 4\log 2$ they remain quite narrow.

The latter implies that a considerable amount of the GW ringdown's modes content cannot be absorbed by a \gls{qbh}. As argued in \cite{Cardoso_2019}, the remaining modes could be reflected, producing a late-time echo in the gravitational waveform. Modeling the reflectivity of QBH is highly non-trivial and requires careful consideration of various (quantum) effects. In this work, we utilize the simplified, phenomenologically motivated toy model described in \cite{Maibach_2024}. The latter includes a broad range of potential echo morphologies and is based on the key assumption that only the characteristic frequencies $\omega_N$ can be absorbed. Including phenomena like line broadening, exponential decay, the characteristic frequencies' sharp lines in the QBH absorption spectrum turn into broader cusp like features with exponential damping. The final reflectivity coefficient for the QBH is found to be \cite{Maibach_2024}
\begin{align}
    \label{equ:reflectivity_QBH}
    \RQBH =e^{-i\omega8\ln{(\beta)}}
     e^{-\frac{|\omega|}{2T_\T{QH}}}\begin{cases} 
            1, & \omega\leq\omega_2 \,, \\
             \left|\sin{\left( \frac{\pi\omega}{\omega_N(\alpha)} \right)}\right|^\delta, & \omega > \omega_1/2 \,.
            \end{cases}
\end{align}
Just as for the reflectivity of the ECO, $\Reco$, equation \eqref{equ:reflectivity_QBH} contains an exponential decay term governed by $T_\T{QH}$. In addition, $\RQBH$ acquires the sine-term with exponent $\delta$ which effectively determines the \textit{sharpness} of the absorption lines corresponding to the characteristic frequencies $\omega_N(\alpha)$. The sine-term itself assures the equidistant distribution of absorption lines\footnote{Note that equidistance is not a universal feature of all quantum gravity theories. For explicit examples and discussions, see \cite{Agullo_2021}.}. The complex exponential, similar to $\Reco$, encapsulates the time shift of the echoes compared to the QNM of the initial GW's ringdown. For the QBH, the exponent is constructed based on the following considerations: The time delay between the onset of the ringdown and the arrival of the first echo at the GW detector is determined by the time duration it takes for the ingoing gravitational radiation - resulting from solving the Teukolsky equation \cite{Teukolsky_1972, Teukolsky_1973} and \textit{morally} emitted from the BH potential barrier\footnote{We will provide a more elaborated picture of the procedure generating GW echoes below.} - to travel to the reflective surface and return to the BH's potential barrier, i.e., 
\begin{align}
\label{equ:time_delay}
    \Delta t_\T{echo} = 2 r_* \big|_{r_\T{Shell}}^{r_\T{Barrier}}\,,
\end{align}
where $r_*= r + 2\ln{(\frac{r}{2}-1)}$ is the tortoise coordinate and $r_\T{Shell}>r_H$ for $r_H = 2M$.
Thereby, the distance between the reflective surface and the BH horizon is crucial for an outside observer's perception due to time dilation GW modes experience close to the horizon. The latter distance is quantified by the parameter $\beta$. For a detailed calculation of equation \eqref{equ:time_delay}, we refer to \cite{Afshordi_I,Wang_2018}.

In summary, the reflectivity $\RQBH$ is determined by four parameters: the phenomenological constant $\alpha$ (which in turn unequivocally determines $\omega_N$), $\beta$, $\delta$, and $T_\T{QH}$. For $\RQBH$, we re-parametrize $T_\T{QH}$ in $\RQBH$ by $T_\T{QH} = \epsilon T_\T{H}$ such that $\epsilon$ steers the exponential decay of the reflectivity coefficient. To better distinguish between the models in the analysis of this work, we keep $T_\T{QH}$ as a free parameter for $\Reco$ such that is completely determined by $\gamma$ and $\epsilon$. It is important to note that the parameters $\gamma$ and $\beta$ only impact the time separation of the echo and the GW in the strain time series picked up by the GW detector. Additionally, the parameter $\alpha$ does not constitute a freely chosen model parameter of reflectivity; rather, it is derived from fundamental properties of the underlying quantum theory of gravity. In Bekenstein's original proposal, the constant takes the value $\alpha=8\pi$ \cite{Bekenstein:1974jk} and subsequent computations relying on different approaches have confirmed his findings \cite{Maggiore_2008}. Nevertheless, alternative values are present in the literature, and we therefore treat $\alpha$ as a free parameter. The actual significance of these model parameters concerning the echo-induced features in the gravitational wave memory will be explored below.

\subsection{Echo generation via the Hybrid Method and Reconstruction of Spacetime}
\label{subsec:echo_reconstruction}
The GW echo is primarily driven by the QNMs associated with the ringdown phase of the GW signal. Despite presenting a significant challenge, various techniques of extracting (analytical) echo templates have been established \cite{Testa_2018,Maggio_2019_II}. In this work, we employ the methodology outlined in \cite{Ma_2022}, where the waveforms near the horizon of an ECO or QBH are reconstructed using asymptotic information contained in the Newman-Penrose scalars $\Psi_0^\circ$ and $\Psi_4^\circ$ at future null infinity, $\scrip$. Below, we summarize the key concepts underlying the reconstruction of GW echoes.

\begin{figure}
	\centering
	\includegraphics[width=0.99\linewidth]{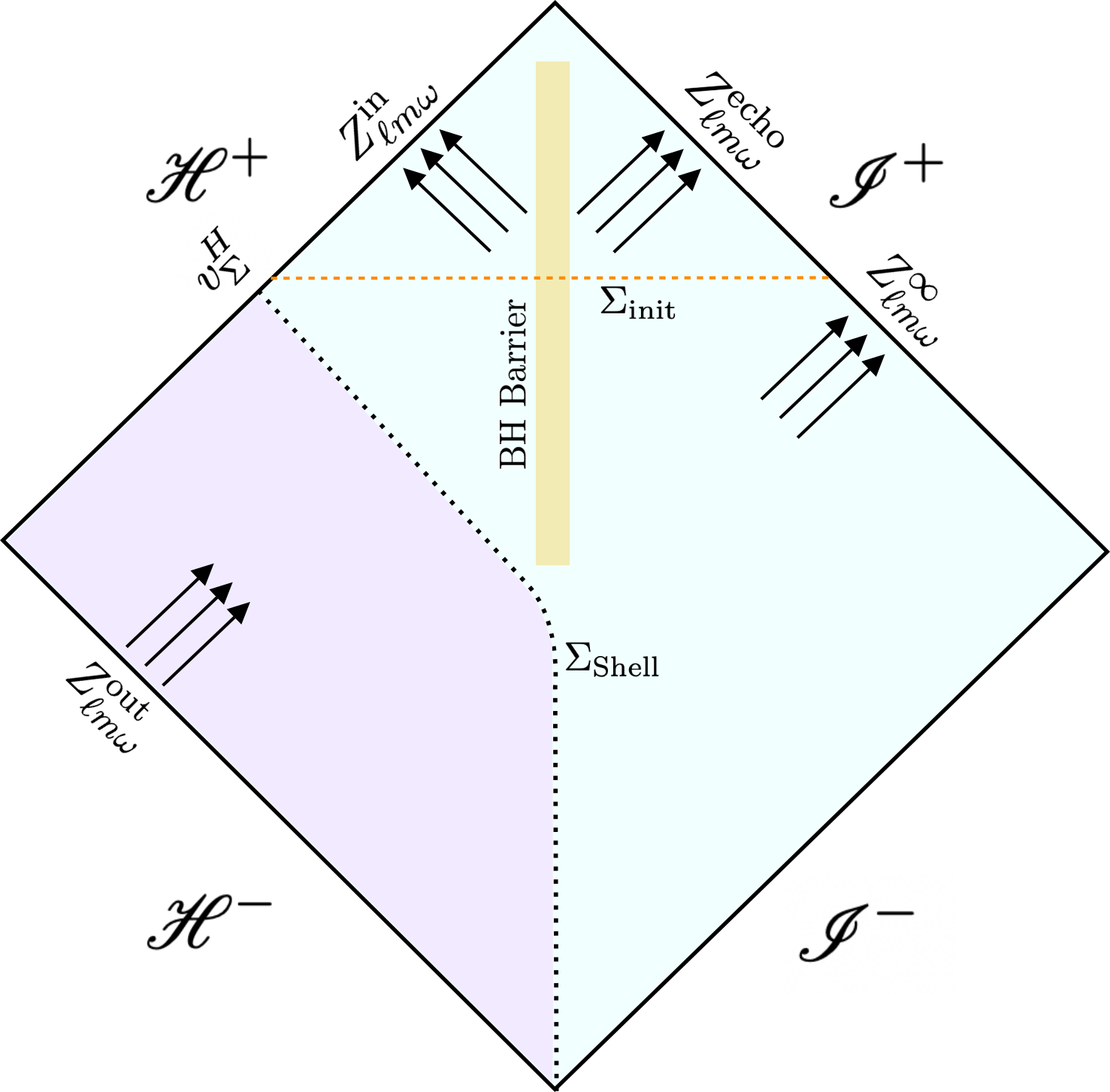}
	\caption{Spacetime diagram illustrating the amplitudes received or emitted by each individual horizon. Here, $\Zinf$ describes the ring down wave while $\Zecho$ denotes all subsequent echoes. The diagram holds analogously also for all $Y_{\ell m \omega}$. The construction outlined in \cite{Ma_2022} considers the blue-shaded regions only. }
	\label{fig:STR}
\end{figure}

The construction of \cite{Ma_2022} largely follows the \textit{hybrid approach} \cite{Nichols_2010, Nichols_2012} in which spacetime is divided into two regions as depicted in Fig. \ref{fig:STR}. For the larger, blue-shaded region, BH perturbation theory is valid, while below the 3-dimensional time-like tube $\Sigma_\T{Shell}$, the system enters the strong-field regime. The exterior region is treated as linearly perturbed Schwarzschild, and for the interior, post-Newtonian (PN) theory is applied, matching the perturbed Schwarzschild metric at $\Sigma_\T{Shell}$\footnote{Naturally, the PN treatment breaks down during the later stages of the BBH evolution but as the shell falls into the future horizon $\Hp$ the errors do not propagate toward $\scrip$.}. For blue-shaded region, the no-incoming wave condition at $\scrim$ is prescribed, while outgoing waves are imposed at $\scrip$. The dynamics in this region are described by the Teukolsky equation. Based on the outgoing gravitational waveform, the perturbative field close to $\Hp$ is determined and subsequently, the echo waveform is derived.

In general, the homogeneous solutions of the Teukolsky equation for $\Psi_0$ and $\Psi_4$ can be decomposed using spin-weighted spheroidal harmonics. By setting the dimensionless spin parameter $a=0$, i.e., considering only mergers with non-spinning remnants, the solutions can be written using spin-weighted spherical harmonics$\,_sY_{\ell m}$, as 
\begin{subequations}
\begin{align}
    \Psi_4(t,r,\theta,\phi) &= \frac{1}{r^4}\sum_{\ell,m} \int \dd \omega \,_{-2}R_{\ell m\omega} \,_{-2}Y_{\ell m }(\theta,\phi)e^{-i\omega t},\\
    \Psi_0(t,r,\theta,\phi) &= \sum_{\ell,m} \int \dd \omega \,_{+2}R_{\ell m\omega} \,_{+2}Y_{\ell m }(\theta,\phi)e^{-i\omega t}.
\end{align}
\end{subequations}
The radial functions $\,_sR_{\ell m \omega}(r)$ can be determined by matching the solutions of the radial Teukolsky equations with solutions determined in BBH coalescence spacetimes, asymptotically given by \cite{Ma_2022}
\begin{subequations}
\label{equ:BH_teukolsky}
    \begin{align}
       \,_{-2}R^{\T{BBH}}_{\ell m \omega} 
       \sim\begin{cases} 
            r^3 Z^\infty_{\ell m \omega}e^{i\omega r^*}, & r^* \rightarrow +\infty, \\
             Z^{\T{out}}_{\ell m \omega} e^{i\omega r^*} + \Delta^2 Z^{\T{in}}_{\ell m \omega}e^{-i\omega r^*}, & r^*\rightarrow - \infty,
            \end{cases}\\
        \,_{+2}R^{\T{BBH}}_{\ell m \omega} 
       \sim\begin{cases} 
            r^{-5}Y^\infty_{\ell m \omega} e^{i\omega r^*}, & r^* \rightarrow +\infty, \\
             Y^{\T{out}}_{\ell m \omega} e^{i\omega r^*} + \Delta^{-2} Y^{\T{in}}_{\ell m \omega}e^{-i\omega r^*}, & r^*\rightarrow - \infty,
            \end{cases}
    \end{align}
\end{subequations}
where $r^*$, again, is the tortoise coordinate. In equation \eqref{equ:BH_teukolsky}, the top solutions approximate the behavior close to $\scrip$, the bottom solutions yield the dynamics close to the BH horizon/ECO surface. A relation connecting the waves escaping to infinity, $\Zinf$ and $\Yinf$ (corresponding to $\Psi_4^\circ$ and $\Psi_0^\circ$, respectively), with the ingoing waves at the future horizon, $\Hp$, $\Zin$ and $\Yin$, is given by \cite{Ma_2022}
\begin{subequations}
\label{equ:coefficients_interpolate}
    \begin{align}
        \Zout  &= \Dout \Zinf\,,\\
        \Zin  &= \Din \Zinf\,,\\
        \Yout  &= \Cout \Yinf\,,\\
        \Yin  &= \Cin \Yinf\,,
    \end{align}
\end{subequations}
where the coefficients $\Din, \Dout, \Cin, \Cout$ are computed using the Black-Hole Perturbation Toolkit \cite{BHPToolkit}. They encapsulate the physics of the BH potential barrier: the transitivity for radiation traveling from $\Hm$ to $\scrip$ is given by $1/\Dout$ ($1/\Cout$), the reflectivity of the potential barrier displayed in Fig. \ref{fig:STR} by $\Din/\Dout$ ($\Cin/\Cout$). Together, the reflective surface of the \gls{qbh} or \gls{eco} and the BH potential barrier form a cavity. The amplitudes in- and outgoing radiation at null infinity, $\Zinf$ and $\Yinf$, can be obtained from numerical relativity (NR) simulations using the  Cauchy-Characteristic-Evolution (CCE) simulation pipeline \cite{Moxon_2020,moxon2021spectrecauchycharacteristicevolutionrapid} incorporated in the numerical relativity code SpECTRE \cite{Kidder_2017,spectrecode}. 

The strain detected by a GW detector, \textit{morally} located at $\scrip$, can be decomposed into two components: the primary waveform of the BBH merger, where we denote its amplitude as $\Zinf$, and subsequent echoes, here denoted as $\Zecho$. The conversion from radial amplitude solution of the Teukolsky equation to GW strain at $\scrip$ is easily written down in Fourier space as
\begin{align}
\label{equ:strain_to_Z}
    h^{\infty}_{\ell m}(\omega) = \frac{1}{\omega^2}\Zinf\,.
\end{align}
For the transformation of ingoing to outgoing solution, we rely on the Teukolsky-Starobinsky (TS) relations 
\begin{subequations}
\label{equ:CD}
\begin{align}
    \frac{4\omega^4}{C^*}\Yinf = \Zinf,\\
    \Yin = \frac{D}{C}\Zin,
\end{align}
\end{subequations}
where
\begin{subequations}
    \begin{align}
        C &= (\ell-1)(\ell +1 )(\ell +2)\ell +12 i\omega,\\
        D &= 64i\omega(128\omega^2 + 8)(1-2i\omega).
    \end{align}
\end{subequations}
An overview of the relations between the asymptotic information at $\scrip$ and at the horizon $\Hp$ for both ingoing and outgoing radiation is provided in Fig. 3 of \cite{Ma_2022}.

For the computation of the echo, the relevant information is encapsulated in the ingoing radiation at the future horizon $\Hp$, here represented by $\Yin$ (see Fig. \ref{fig:STR}). This ingoing radiation travels towards the (partially) reflective surface, where part of it trapped in the cavity. To determine how each frequency mode interacts with the BH or ECO, both a reflectivity parameter and an appropriate boundary condition are required. With the reflectivity coefficients already established in the previous subsection, we now turn to the selection of a suitable boundary condition: For the ECO, we apply the same approach as in \cite{Ma_2022} relying on an earlier work considering the tidal forces applied on zero-angular-momentum fiducial observers \cite{Chen_2021}. By computing the transverse component of the tidal tensor field from the point of view of these fiducial observers, the ECO's boundary condition can be derived as
\begin{align}
\label{equ:boundary_ECO}
    \Zouteco{}^{,1} = \frac{(-1)^{\ell + m +1 }}{4} \Reco \Yineco \,,
\end{align}
where the upper index on the left-hand side signifies that equation \eqref{equ:boundary_ECO} computes the first echo only (one for each cycle the radiation undergoes within the cavity). $\Yineco$ is determined by the ingoing $\Psi_0$ which takes the form of curvature perturbations of fiducial observers near the horizon and includes only the fraction of $\Yin$ associated with the QNM of the ringdown phase\footnote{As $\Yin$ is inferred from the full waveform at $\scrip$, $\Yinf$, via \eqref{equ:coefficients_interpolate}, it naturally carries information about all regimes of the waveform. However, physically, only the ringdown part exists within the potential barrier and can fall down the horizon $\Hp$. The procedure rectifying this issue is explained below.}. In the observer's frame, the ECO gives rise to a local reflectivity $\Reco$ and an additional conversion factor of $(-1)^{\ell + m +1 }/4$. For the outgoing amplitude defined in \eqref{equ:boundary_ECO} to reach $\scrip$, this first echo has to be transmitted through the BH barrier and, thus,
\begin{align}
     \Zecho{}_{,1} = \frac{1}{ \Dout} \Zouteco{}^{,1}  \,.
\end{align}
A similar procedure can be applied to the QBH reflectivity model, for which the boundary condition has not yet been specified. From a phenomenological standpoint, there is no compelling reason to assume significant alterations to the ingoing wave beyond the possibility of a phase shift. All other effects are encapsulated in the model-dependent reflectivity coefficient $\RQBH$. Therefore, inspired by the arguments that lead to equation \eqref{equ:boundary_ECO}, we postulate the boundary condition for the QBH to read \footnote{We acknowledge that the boundary condition in \eqref{equ:boundary_ECO} is fundamentally tied to the tidal deformability of ECOs. As classical BHs do not exhibit tidal deformability, the analogy is lacking here. Consequently, it is possible that BH could respond in a completely different manner. However, for the purposes of this study, we adopt a slightly generalized ECO-like boundary condition and leave the intriguing question of the appropriate QBH boundary condition to future investigations.}
\begin{align}
\label{equ:boundary_QBH}
    \Zoutqbh{}^{,1} = \frac{(-1)^{\ell + m +1 }}{\zeta} \RQBH \Yinqbh \,.
\end{align}

As mentioned above, the boundary conditions \eqref{equ:boundary_ECO} and \eqref{equ:boundary_QBH} represent only the first echo of the GW. However, as the potential barrier is not perfectly transmitting, a portion of the radiation $\Zouteco$ ($\Zoutqbh$) is reflected back (with the reflectivity given by $\Din/\Dout$) and the cycle starts over (see Fig. \ref{fig:cavity}). The reflected quantity can again be transformed into ingoing radiation via the relations above. For simplicity, we include all necessary transformations as well as the numerical factor of $(-1)^{\ell+m+1}$ into 
\begin{align}
\label{equ:RBH}
    \RBH = (-1)^{\ell+ m +1 } \frac{\Din}{\Dout}\frac{D}{4C}\,,
\end{align}
representing the reflectivity of the potential barrier. Using the latter, we can compute the full echo as
\begin{align}
\label{equ:sum_echo}
    \Zecho &= \frac{(-1)^{\ell +m+1}\Reco}{1-\Reco\RBH}\frac{1}{4\Dout}\Yineco\notag\\
    &= \frac{C}{D \Din}\sum_{n=1}\left(\Reco\RBH\right)^n \Yineco\notag\\
    &=\sum_n\Zecho{}_{,n}\,.
\end{align}
In the latter equation, the total echo is rewritten as a sum over individual contributions, each associated with $n$ cycles in the cavity formed by the barrier and the reflective surface. 

\begin{figure}
	\centering
	\includegraphics[width=1.1\linewidth]{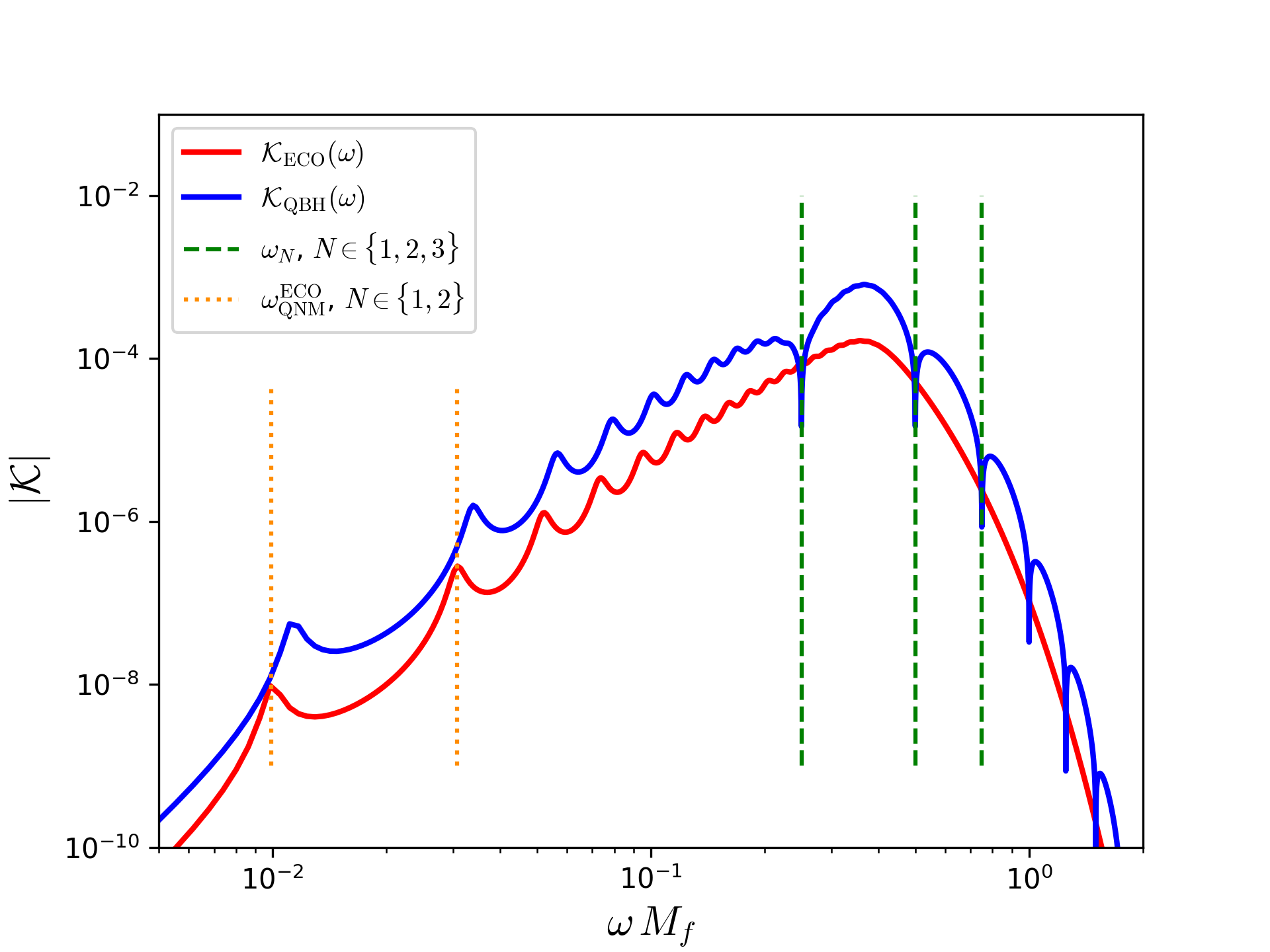}
	\caption{Transfer functions for $\Reco$ and $\RQBH$ with $T_\T{QH}= 1/8\pi, \gamma = 10^{-15}$ and $\alpha = 8\pi, \delta = 0.5, \epsilon = 1, \beta = 10^{-15}$, $\zeta=4$. The green dashed lines mark the characteristic absorption frequencies for the QBH. The dotted orange lines mark the QNM of the ECO. Similarly, the peaks of $\mathcal{K}_\T{QBH}$ mark the QNMs of the QBH. For better readability, the transfer function of the QBH, $\mathcal{K}_\T{QBH}$, is multiplied with an overall factor of 5.}
	\label{fig:Transfer_functs}
\end{figure}

To avoid instabilities of the ECO, one must ensure that $|\Reco\RBH|<1$ at all times. An exception to the inequality are the QNM, $\omega_n$, of the ECO (QBH) for which $\Reco(\omega_n)\RBH(\omega_n)=1$ ($\RQBH(\omega_n)\RBH(\omega_n)=1$). The QNMs therefore appear as poles of the corresponding transfer function 
\begin{align}
    \mathcal{K}(\omega) := \frac{C}{D \Din}\sum_{n=1}\left(\Reco\RBH\right)^n \,.
\end{align}
The transfer functions for $\Reco$ and $\RQBH$ are displayed in Fig. \ref{fig:Transfer_functs}.
We clearly see the difference in the QNMs for ECO and QBH. The global maxima, however, align as they both represent the fundamental QNM frequency of the Schwarzschild BH. The roots of the QBH's reflectivity are indicated by dashed green lines. For large frequencies, both transfer functions are suppressed exponentially. For small frequencies, the $1/\Dout$ factor in equation \eqref{equ:sum_echo} dominates the transfer function. For the ECO, both $T_\T{QH}$ and $\gamma$ determine its QNMs. Additionally, $T_\T{QH}$ determines the exponential suppression. For the QBH, $\beta, \delta$ impact the location of the QBH's QNMs, $\omega_\T{QNM}^\T{QBH}$, while $\alpha$ shifts the roots and $\epsilon$ governs the exponential decay. The overall amplitude is suppressed by $\zeta$. All parameters and their correspondence between reflectivity models are summarized in Table \ref{table:1}.
\begin{center}
\begin{table*}[t]
\begin{tabular}{
    |c|c|c|
}
 \hline \bfseries
 $\mathcal{K}(\omega)$ attribute & \bfseries $\,\,\,\,$Parameter ECO$\,\,\,\,$ & \bfseries $\,\,\,\,$Parameter QBH $\,\,\,\,$
 \\\hline\hline 
 Time separation echo & $\gamma, T_\T{QH}$   & $\beta$   
 \\\hline
  $\,\,\,\,$ Exponential supp. for large $\omega$ $\,\,\,\,$& $T_\T{QH}$&   $\epsilon$  
   \\\hline
  Location QNMs & $\gamma, T_\T{QH}$  & $\beta,\epsilon$ 
  \\\hline
Separation of roots & - & $\alpha$ 
  \\\hline
 ``Sharpness'' of roots & -  & $\delta$ 
 \\\hline
 Boundary suppression & 1/4  & 1/$\zeta$ 
 \\\hline
\end{tabular}
 \caption{
 Model parameters for the reflectivity of ECOs and QBHs. An implementation of the transfer functions $\mathcal{K}(\omega)$ is displayed in Fig \ref{fig:Transfer_functs}. As the ECO's transfer function does not obtain roots, there is no correspondence between $\alpha, \delta$ for QBHs.
    }
    \label{table:1} 
\end{table*}
\end{center}
The factors $|\Reco|^2$ and $|\RQBH|^2$ represent the corresponding energy reflectivity of the ECO and QBH, respectively. Similarly, $|\RBH|^2$ represents the energy reflectivity of the potential barrier \cite{Xin_2021}. Based on this interpretation, we can equally define a coefficient of transmission for the amplitude of ingoing radiation at $\Hp$. We define, for ECO and QBH analogously, 
\begin{align}
\label{equ:energy_transmittivity}
    |\Teco|^2 := 1- | \Reco|^2 \,,
\end{align}
such that, for the initial amplitude penetrating the ECO's surface (the QBH's reflective shell) and propagating towards the horizon $\Hp$, we find
\begin{align}
    \ZBH{}^{,1} = \Teco\Yineco\,.
\end{align}
As for the reflected radiation sourcing the echo in equation \eqref{equ:sum_echo}, the transmitted portion receives a contribution for every cycle. Adding all subsequently transmitted radiation, we obtain
\begin{align}
\label{equ:sum_echo_BH}
    \ZBH &=  \left(\Teco+ \frac{\Teco\Reco\RBH}{1-\Reco\RBH}\right)\Yineco\notag\\
    &= \Teco  \sum_{n=1} \left(\Reco\RBH\right)^{n-1}\Yineco\notag\\
    &=\sum_n \ZBH{}^{,n} \,.
\end{align}
Equations \eqref{equ:sum_echo} and \eqref{equ:sum_echo_BH} will be used in subsequent sections to determine the energy and angular momentum flux across both horizons $\scrip$ and $\Hp$. We stress at this point that equation \eqref{equ:energy_transmittivity} does not fully determine $\Teco$ as a complex phase remains undetermined. Technically, this issue can only be resolved by adding further assumptions to the model. However, in this work, we are solely interested in the energy and angular momentum fluxes. As equation \eqref{equ:energy_transmittivity} precisely describes the energy transmitivity, its definition is sufficient for the subsequent analysis.

Both the amplitude falling into the BH, $\ZBH$, and the one escaping to $\scrip$, $\Zinf$, are determined by $\Yineco$, the ingoing radial component of the $\Psi_0$-wave falling down the future horizon in the classical picture. Following \cite{Ma_2022}, $\Yineco$ can be determined by computing the late-time portion of $\Yin$ that corresponds to the QNM, i.e., the ringdown phase of the full waveform. The relevant portion is determined by the retarded time parameter, $v^H_\Sigma$, associated to the shell $\Sigma_{\T{shell}}$ separating the two spacetime regions in Fig. \ref{fig:STR}. With $v^H_\Sigma$ at hand, one can define \cite{Ma_2022}
\begin{align}
\label{equ:filter_Y}
    Y^\T{in ECO}_{\ell m} (v) = Y^\T{in}_{\ell m} (v) \mathcal{F}(v) + \T{Const.} \cdot \big(1-\mathcal{F}(v)\big)\,,
\end{align}
with $\mathcal{F}(v)$ being the Planck-Taper filter as defined in \cite{Ma_2022}. Physically speaking, the Planck-Taper filter guarantees the desired inclusion of only the ringdown-affiliated part of the waveform. The shell time $v^H_\Sigma$ is determined via minimizing the NR waveforms mismatch with a sample waveform purely constructed from QNM overtones with different initial times (see \cite{Ma_2022} for details). 

With the algorithm sketched above, determining the final echo-corrected waveform numerically requires solely the NR data for $\Psi_0^\circ$ and $\Psi_4^\circ$ defined at $\scrip$. As mentioned above, we obtain the NR data through CCE simulations. The latter adopts the conventional normalization, that is,
\begin{subequations}
\label{equ:NR_data}
    \begin{align}
        rM_f\Psi_4^\circ = \sum_{\ell, m} \,_{-2}Y_{\ell m}(\theta,\phi)Z^\infty_{\ell m } \,,\\
        rM_f^{-1}h^\circ = \sum_{\ell, m} \,_{-2}Y_{\ell m}(\theta,\phi)\hnorm_{\ell m} \,, \\
        r^{5}M_f^{-3}\Psi_0^\circ = \sum_{\ell, m} \,_{+2}Y_{\ell m}(\theta,\phi)Y^\infty_{\ell m } \,.
    \end{align}
\end{subequations}
Here, $M_f$ is the mass of the remnant compact object. 

\begin{figure}
\includegraphics[width=1.0\columnwidth]{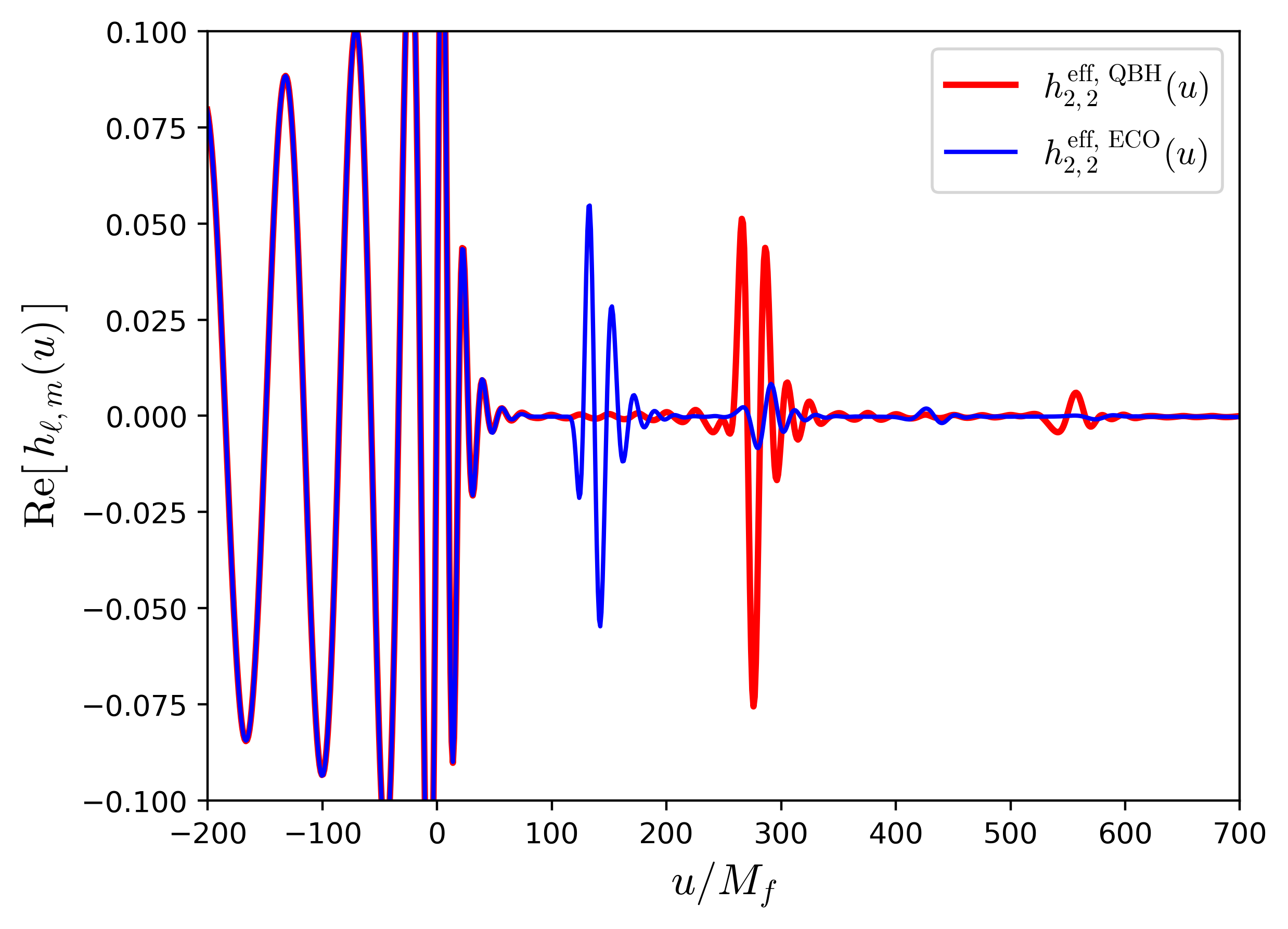} 
\includegraphics[width=1.0\columnwidth]{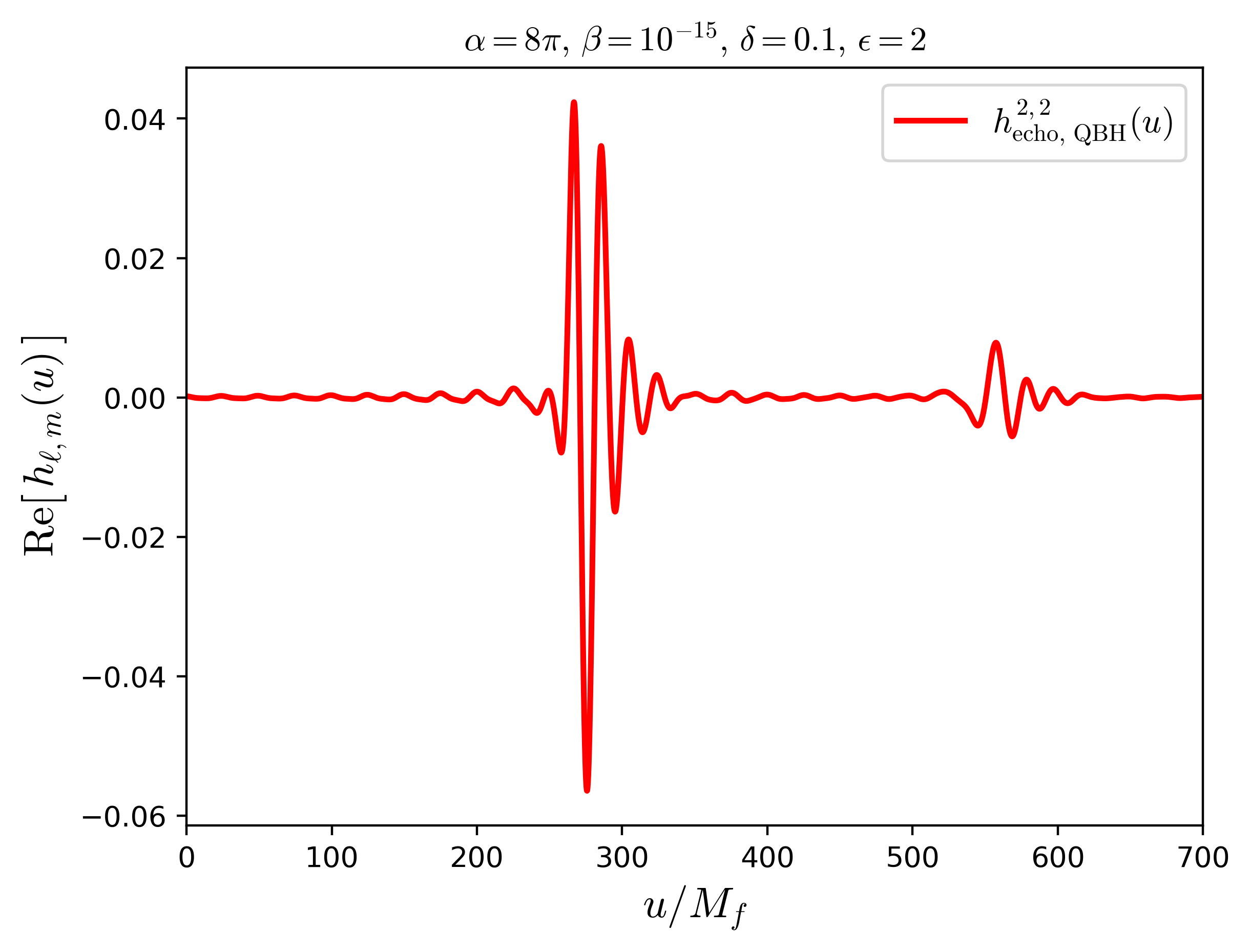}   
     \caption{Full waveform $\hfull=\hecho+\hnorm$ (top) for ECO and QBH simulated for event \textit{SXS:BBH:0207}. For the ECO, we choose $T_\T{QH}=2T_\T{H}$ and $\gamma=10^{-15}$. For the QBH, the parameter choice is displayed on top of the bottom plot. The latter displays the isolated $\hecho$ from the upper plot.}
       \label{fig:waveform_QBH}
\end{figure}

Equation \eqref{equ:sum_echo} provides an expression for the strain of the GW echo, determined by a given reflectivity model and the properties of the original GW signal at $\scrip$, up to a frequency-dependent prefactor. At this stage, it is worth mentioning that this contribution adds on top of the original waveform. As these signals and the process of origin are disentangled, one can simply compute the complete waveform as
\begin{align}
\label{equ:strain_decomposition}
    \hfull = \hecho + \hnorm\,,
\end{align}
where $\hnorm$ results from $\Zinf$, the \textit{raw} asymptotic waveform at $\scrip$. A depiction of $\hfull$ is provided in Fig. \ref{fig:waveform_QBH}. Thereby, $\hecho$ can be explicitly computed as a convolution of all aforementioned sub-steps, using equations \eqref{equ:coefficients_interpolate}, \eqref{equ:strain_to_Z}, \eqref{equ:sum_echo}, \eqref{equ:filter_Y}, and \eqref{equ:NR_data} as
\begin{align}
\label{equ:hecho}
    \hecho(\omega) &= \sum_n \sum_{\ell, m} \,_{-2}Y_{\ell m}(\theta, \phi)\mathfrak \hecho_{\ell m, n}(\omega)\,,
\end{align} 
where 
\begin{align}
\label{equ:some_label}
\hecho_{\ell m, n}(\omega)=&\frac{1}{\omega^2}\frac{C}{D \Din}\left(\Reco\RBH\right)^n \notag\\&\cdot\mathfrak F\left\{C^\T{in}_{\ell m} \Psi_{0, \ell m}(v)\mathcal{F}(v) \right\}\,,
\end{align}
and we define $\Psi_{0, \ell m }$ such that
\begin{align}
    r^5M_f^{-3}\Psi_0^\circ = \sum_{\ell,m} \,_{+2}Y_{\ell m}(\theta, \phi) \Psi_{0,\ell m}\,,
\end{align}
where $\mathfrak F (\Psi_{0,\ell m}) =: \Yinf$ and $\mathfrak F (\cdot)$ denotes the Fourier transform. Note also that $C^\T{in}_{\ell m}$ represents the Fourier transform of $\Cin$. Exemplary echoes are displayed in Fig. \ref{fig:waveform_QBH}. The parameter dependence of QBH echoes is displayed in Fig. \ref{fig:echo_QBH} of Appendix \ref{app:A} (compare with Table \ref{table:1}). A similar plot for ECOs is found in \cite{Ma_2022}. As expected, the two types of echoes exhibit only slight variations in their morphology.

The statement in equation \eqref{equ:strain_decomposition} is independent of the time separation that is determined by the corresponding parameters of a given reflectivity model, i.e., independent of the complex phase factors in \eqref{equ:reflectivity} and \eqref{equ:reflectivity_QBH}. Therefore, the echo can easily be added to (numerical) waveforms by adding it on top of the raw strain time series in a post-processing step similar to \cite{Mitman_2021}. Crucially, the echo, as well as its memory, can be treated as individual ``events''\footnote{This approach aligns with recent proposals suggesting that echoes should be treated as independent signals rather than as extensions of the BBH merger waveform due to the potentially substantial time delays involved \cite{zimmerman2023rogueechoesexoticcompact}.}.

\section{Memory and Flux for (Quantum) Gravitational Waveforms}
\label{sec:balance_laws}

In the literature, numerous studies have explored the GW memory based on distinct approaches \cite{Bieri_mem, Favata_mem, Throne_mem, Blanchet_mem, Zeldovich_mem, Christodoulou_mem}. The equivalence of the resulting expressions for memory has been well established. For applications involving numerical waveforms, a specific formulation of the gravitational memory based on the so-called energy-momentum or Bondi-Metzner-Sachs (BMS) balance laws \cite{Ashtekar_2019} (see also \cite{heisenberg2023balancelawstestgravitational, DAmbrosio_2022} for more technical details) has become widely adopted \cite{Mitman_2020, Khera_2021, Mitman_2021}. These balance laws provide a series of constraint equations that relate the GW strain at future null infinity to physical properties of the remnant compact object, such as its velocity and mass. Notably, they include expressions for both the linear and non-linear GW memory. These flux balance laws have been applied in prior studies for waveform analysis and NR applications, among others \cite{DAmbrosio_2024, Borchers_2023, Borchers:2021vyw}.

\subsubsection{Computing the Memory}

For the numerical analysis in this work, we utilize the memory components derived from the BMS balance laws as implemented in \cite{Mitman_2021}. This choice is driven by the applicability of these equations to NR waveforms and the convenient format in which they are presented. Specifically, the memory components, as outlined in \cite{Mitman_2021}, enable direct computation of time-dependent memory corrections for each individual harmonic strain mode. Since the BMS laws are derived from full, non-linear GR, we expect these equations to apply equally well to any strain (quantum) corrections that arise within the regime of validity of GR as an effective theory.

The memory equation in \cite{Mitman_2021} is directly derived from the balance flux laws as described in \cite{Ashtekar_2019}. The time series containing only the memory-induced contributions to the strain, after a few simplifications, can be written as
\begin{align}\label{equ:strain_mem}
    \hmem = \frac{1}{2} \bar{\eth}^2 \mathfrak{D}^{-1}\left[\frac{1}{4}\int_{u_i}^u \dd u |\dot h|^2 - \left(\Psi_2 + \frac{1}{4}\dot h \bar h\right)\right]\,,
\end{align}
where $\mathfrak{D}^{-1}$ is defined via
\begin{align}
    \mathfrak D = \frac{1}{8} D^2(D^2 + 2)\,,\\
    D^2 = \bar \eth \eth\,,
\end{align}
and $\eth$ defines the spin-weighted derivative operator whose precise definition can be found in~\cite{Ashtekar_2019, Mitman_2021, DAmbrosio_2022}. The $h$ appearing in equation \eqref{equ:strain_mem} is the full GW strain. It is fundamentally dependent on the line-of-sight, spanning from the detector to the binary frame. The latter can be characterized by two angles $\theta, \phi$, marking a point on the celestial sphere. The initial time $u_i$ in equation \eqref{equ:strain_mem} marks the beginning of the gravitational waveform. As, in this work, we consider numerical waveforms, by construction, we do not use the precise definition of equation \eqref{equ:strain_mem} in which one would take the limit $u_i\rightarrow -\infty$ \cite{Mitman_2021} but instead identify $u_i$ with the starting time of the provided NR waveforms. This truncation must be compensated by an angular dependent constant $\alpha$\footnote{Additionally, frame related issues might contribute to constant angular-dependent terms \cite{Mitman_2021}. Here, we will summarize both contributions into a single term.}. 

The full memory given by equation \eqref{equ:strain_mem} can be separated into a dominating term associated with the energy flux and an oscillatory term, 
\begin{subequations}
\label{equ:mem_gross}
\begin{align}
    h_{\mathcal E} = \frac{1}{2} \bar \eth^2 \mathfrak D^{-1} \left[ \frac{1}{4} \int_{u_i}^u \dd u |\dot h|^2 \right] + \alpha\,,\\
    h_{\T{osc.}} = \frac{1}{2} \bar \eth^2 \mathfrak D^{-1} \left[- \left(\Psi_2 + \frac{1}{4} \dot h\bar h\right) \right]\,,
\end{align}
\end{subequations}
respectively. For the purposes of this work, we further decompose the latter terms such that each component allows for more phenomenological interpretation. Following \cite{Mitman_2020}, equation \eqref{equ:mem_gross} (but in particular $h_{\T{osc.}}$) can be rewritten as
\begin{subequations}
\label{equ:mem_net}
    \begin{align}
        h_m &= \frac{1}{2} \bar \eth^2 \mathfrak D^{-1} m\,,\\
        h_\mathcal{E} &= \frac{1}{2} \bar \eth^2 \mathfrak D^{-1} \left[ \frac{1}{4} \int_{u_i}^u \dd u |\dot h|^2 \right] + \alpha\,,\\
        h_{N} &=\frac{1}{2} i\bar \eth^2 \mathfrak D^{-1} D^{-2} \T{Im}\left\{\bar \eth (\partial_u N)\right\}\,,\\
        h_\mathcal{J} &=\frac{1}{2} i\bar \eth^2 \mathfrak D^{-1} D^{-2} \T{Im}\left\{ \frac{1}{8}\eth(3h\bar \eth \dot{\bar{h}} - 3 \dot h \bar \eth \bar h + \dot{\bar{h}}\bar \eth h- \bar h \bar \eth \dot h)\right\} \,.
    \end{align}
\end{subequations}
Here, $m$ and $N$ correspond to the Bondi mass aspect and the angular momentum aspect respectively \cite{Mitman_2020}, i.e., in terms of strain
\begin{subequations} and Newman Penrose scalar they read
    \begin{align}
        m &= -\T{Re}\left\{\Psi_2 + \frac{1}{4}\dot h\bar h\right\}\,\\
        N &= 2 \Psi_1 - \frac{1}{4} \bar h \eth h - u \eth m - \frac{1}{8}\eth(h\bar h)\,.
    \end{align}
\end{subequations}
While, in principle, we can compute the deviations due to echoes present in the strain time series for all components given in \eqref{equ:mem_net}, we are mainly interested in the components associated with the energy and angular momentum flux across the horizon $\scrip$, i.e., $h_\mathcal{E}$ and $h_\mathcal{J}$. Both components can be related to the \textit{null memory} (also called \textit{non-linear memory}) but have different parity, i.e., $h_\mathcal{E}$ is electric and $h_\mathcal{J}$ is magnetic. For practical purposes, it is even sufficient to consider solely $h_\mathcal{E}$ as it largely dominates the non-linear memory. Being interested in the kinematics of the considered binary, we nevertheless include the computation of the memory associated with the angular momentum flux.

Applying the notation of Section \ref{subsec:echo_reconstruction}, we can identify and separate contributions in $h_\mathcal{E}$ and $h_\mathcal{J}$. We start by considering $h_\mathcal{E}$, where equation \eqref{equ:strain_decomposition} leads to mixed terms of the form
\begin{align}
    \dot h ^{\T{eff}} \dot{\bar{h}}^\T{eff} = |\dot{h}^\infty|^2 + \dot{h}^\infty\dot{\bar{h}}^\T{echo} + \dot{\bar{h}}^\infty\dot{h}^\T{echo} + |\dot{h}^\T{echo}|^2.
\end{align}
Both $\hecho$ and $\hnorm$ being time series waveforms, we assume that the echo, and thus also its time derivative, is time-wise clearly separated from the main waveform $\hnorm$ as, for instance, depicted in Fig. \ref{fig:waveform_QBH}. In this case, we find $\dot{h}^\infty\dot{\bar{h}}^\T{echo} = \dot{\bar{h}}^\infty\dot{h}^\T{echo} = 0$ and the memory $h_\mathcal{E}$ separates into a waveform and an echo-contribution. The assumption of time-wise separability requires a sufficiently large phase shift incorporated in the reflectivities $\Reco, \RQBH$. The relevant parameter bounds can be quantified as $\gamma\lesssim 10^{-4}$ (for $T_\T{QH} = T_\T{H}$, and $T_\T{QH} \lesssim 5 T_\T{H}$ for $\gamma=10^{-15}$) for the echo of an ECO, and for echoes induced by QBH's one finds $\beta\lesssim 10^{-7} $. For the remainder of this work, we restrict ourselves to a range in parameter space where this no-overlap assumption holds valid, such that
\begin{align}
\label{equ:mem_electic_echo}
    \hfull_{\mathcal{E}} &= \frac{1}{2} \bar \eth^2 \mathfrak D^{-1} \left[ \frac{1}{4} \int_{u_i}^u \dd u |\dot h^\T{echo}|^2  + \frac{1}{4} \int_{u_i}^u \dd u |\dot h^\infty|^2 \right]\notag\\
    &=: \hecho_\mathcal{E} + \hnorm_\mathcal{E}\,,
\end{align}
where we omit the constant $\alpha$.

\begin{figure}[t]
\includegraphics[width=1.0\columnwidth]{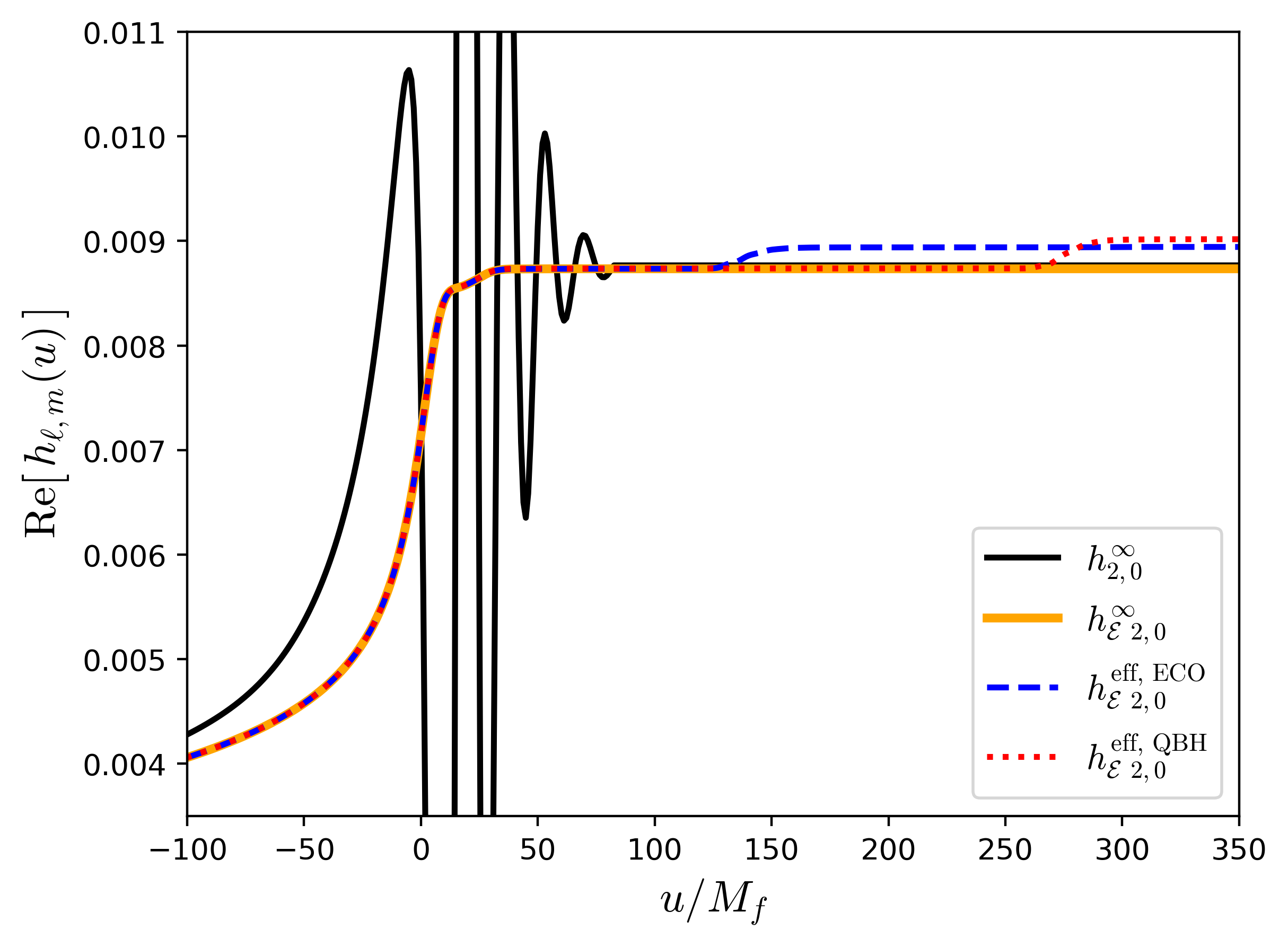} 
\includegraphics[width=1.0\columnwidth]{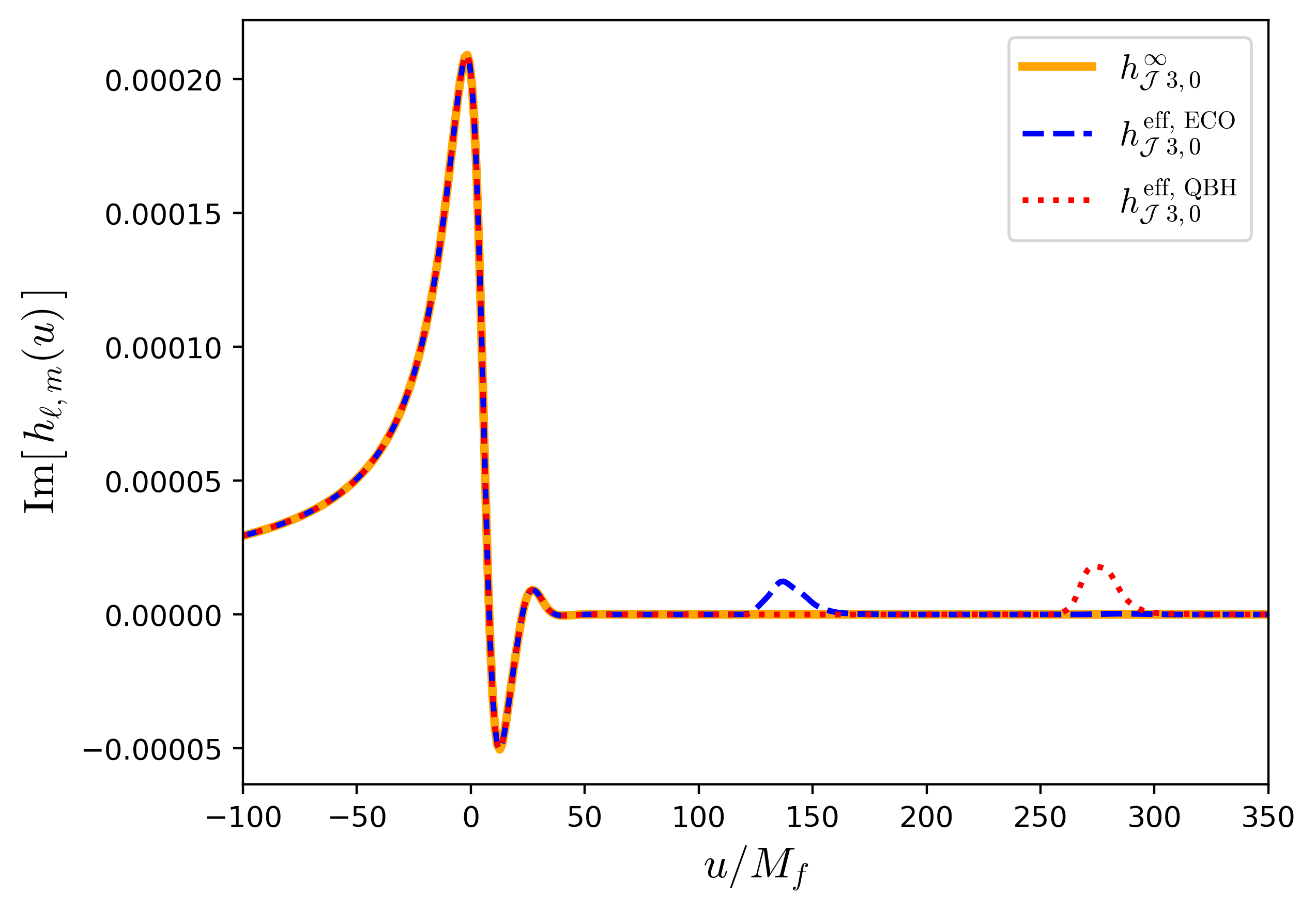}   
     \caption{Selected modes of the memory corrected waveform. The dashed and dotted lines mark the memory contributions $h_\mathcal{E}$ and $h_\mathcal{J}$ from ECO and QBH directly related to the echoes. The ``no-echo'' counterpart is displayed in both the bottom and top plots as the solid orange line. The parameter choice and the event are copied from Fig. \ref{fig:waveform_QBH}.}
       \label{fig:mem_echo}
\end{figure}

Next, we consider $h_\mathcal{J}$. Structurally, the main difference with respect to $h_\mathcal{E}$ are the angular derivative $\eth h$. We argue that the derivative operator only acts on the angular dependent part of the strain, i.e., we can decompose
\begin{align}
    h(u,\theta,\phi) = \sum_{\ell,m } h_{\ell m}(u) \,_{-2}Y_{\ell m}(\theta, \phi)\,,
\end{align}
with
\begin{align}
    \eth\,_sY_{\ell m} = \sqrt{(\ell -s)(\ell +s+1)}\,_{s+1}Y_{\ell m }\,\\
    \bar \eth\,_sY_{\ell m} = - \sqrt{(\ell +s)(\ell -s+1)}\,_{s-1}Y_{\ell m }\,,
\end{align}
such that the strain modes, and thus $\Zin$ ($\Yin$), as functions of retarded (advanced) time, are not affected. Consequently, even with the angular derivative operator $\eth$ applied on the strain, the separation argument still holds, and terms like $\eth \hnorm \bar \eth \hecho$ vanish due to the absence of non-trivial time series overlap. Thus, inserting \eqref{equ:strain_decomposition} in $h_\mathcal{J}$ we find that $\hecho$ does not mix with $\hnorm$ and thus 
\begin{align}
    \hfull_\mathcal{J} = \hecho_\mathcal{J} + \hnorm_\mathcal{J}\,,
\end{align}
where $\hecho_\mathcal{J}$ is simply
\begin{align}
\label{equ:mem_angular_echo}
    \hecho_\mathcal{J} &=\frac{1}{16} i\bar \eth^2 \mathfrak D^{-1} D^{-2} \T{Im}\left\{\eth(3h^\T{echo}\bar \eth \dot{\bar{h}}^\T{echo}\right.\notag\\
    &\left.- 3 \dot h^\T{echo} \bar \eth \bar h^\T{echo}+ \dot{\bar{h}}^\T{echo}\bar \eth h ^\T{echo}- \bar h^\T{echo} \bar \eth \dot h^\T{echo})\right\}\,.
\end{align}
A comparison between $\hecho_\mathcal{J}$, $\hecho_\mathcal{E}$ and $\hnorm_\mathcal{J}$ and $\hnorm_\mathcal{E}$ is displayed in Fig. \ref{fig:mem_echo}. In this case, the time separation of the effects is clearly exhibited. Note that while for $h_\mathcal{E}$ the contributions of waveform and echo are identical in shape (besides the amplitude), for $h_\mathcal{J}$ they are fundamentally different. Note also that if the no-overlap assumption is violated, the step-like shape of the non-linear memory receives non-negligible corrections from the mixing terms $\dot{\bar{h}}^\infty\dot{h}^\T{echo} + c.c.$ which are generally of oscillatory nature. The latter is a fundamental manifestation of the memory's non-linear nature. Thus, a echo could lead to a significant alteration of the ``known'' memory contribution associated to BBH waveforms.

\subsubsection{Computing Fluxes}
The balance flux laws not only provide a framework to incorporate the memory into the GW strain, but they also yield explicit fluxes of physical properties, such as energy and angular momentum, that pass through a detector placed at $\scrip$ or fall into the \gls{bh}. Using equations \eqref{equ:mem_net}, we can extract both the energy and angular momentum fluxes carried by the GWs. Since these equations are separable into components corresponding to the initial strain and the echo-related contributions, so are the flux formulas. Concretely, for a generic strain, the (dimensionless) energy and momentum flux per unit time and angle read \cite{Mitman_2020}
\begin{subequations}
\label{equ:fluxes}
\begin{align}
    \frac{\dd E}{\dd \Omega \dd u} &= \frac{1}{16\pi} {|\dot h|^2} \,,\\
    \frac{\dd J}{\dd \Omega \dd u} &= \frac{1}{16\pi} \left(3\bar h \eth \dot{h} - 3 \dot{\bar{h}} \eth  h + \dot{h} \eth \bar h -  h \eth \dot{\bar{h}}\right)\,.
\end{align}
\end{subequations}
The flux formulas \eqref{equ:fluxes} are defined for asymptotic strains at $\scrip$, hence the choice of the time coordinate, $u$.
Strictly speaking, in equations \eqref{equ:fluxes} (and all memory definitions before), one should replace strain $h$ with the asymptotic strain $h^\circ$ defined exclusively at $\scrip$, and which continues into physical spacetime via
\begin{align}
\label{equ:asymptotic_strain}
    h_+^\circ = \lim_{r\rightarrow \infty} rh_+, && h_\times^\circ = \lim_{r\rightarrow \infty} rh_\times\,.
\end{align}
The aforementioned definition becomes especially important when computing the dimensionful versions of \eqref{equ:fluxes}. In this case, the asymptotic strain must be replaced by the physical strain. For practical examples, such as actual measurements of \gls{bbh} mergers, the radius is replaced by the estimated luminosity distance $D_L$ of the event. Naturally, in addition to the distance information, appropriate powers of $c$ and $G$ have to be factored in to maintain dimensional consistency. 

When computing \eqref{equ:fluxes}, it may be helpful to make use of strain's decomposition into spherical harmonics outlined, for instance, in \cite{DAmbrosio_2024} (see also Appendix \ref{app:decomp}). For instance, for the memory flux associated with the echo one computes
\begin{align}
    \frac{\dd E^\T{echo}}{\dd \Omega \dd u} &= \frac{1}{16\pi} |\dot{h}^\T{echo}|^2\,,
\end{align}
which can be decomposed into spherical harmonics of spin-weight zero,
\begin{align}
     |\dot{h}^\T{echo}|^2 = \sum_{\ell, m} \alpha^\T{echo}_{\ell m} Y_{\ell m}(\theta, \phi)\,.
\end{align}
The resulting frequency-dependent expansion coefficients, up to a prefactor depending on $\ell$ and $m$ (see Appendix \ref{app:decomp} for details), read 
\begin{widetext}
   \begin{align}
     \alpha^\T{echo}_{\ell m \omega} \sim \sum_n\sum_m\frac{1}{\omega^4}\left(\frac{C}{D}\right)^2(\Reco\RBH)^{n+m}\frac{1}{D^\T{in}_{\ell_1 m_1 \omega}D^\T{in}_{\ell_2 m_2 \omega}}\partial_v\mathfrak F\left[C^\T{in}_{\ell_1 m_1} \Psi_{0, \ell_1 m_1}(v)\mathcal{F}(v)\right]\partial_v\mathfrak F\left[\overline{C^\T{in}_{\ell_2 m_2} \Psi_{0, \ell_2 m_2}(v)\mathcal{F}(v)}\right]\,.
\end{align} 
\end{widetext}
A similar result holds for the angular momentum flux ${J}^\T{echo}$, expressed in terms of coefficients $\beta^\T{echo}_{\ell m \omega}$. The coefficients $\alpha^\T{echo}$ and $\beta^\T{echo}$ differ solely by a mode-dependent factor, the explicit forms of which are provided in Appendix \ref{app:decomp}. It is important to note that this decomposition applies equally to the memory formulas \eqref{equ:mem_electic_echo} and \eqref{equ:mem_angular_echo}.

\begin{figure}
	\centering
	\includegraphics[width=0.99\linewidth]{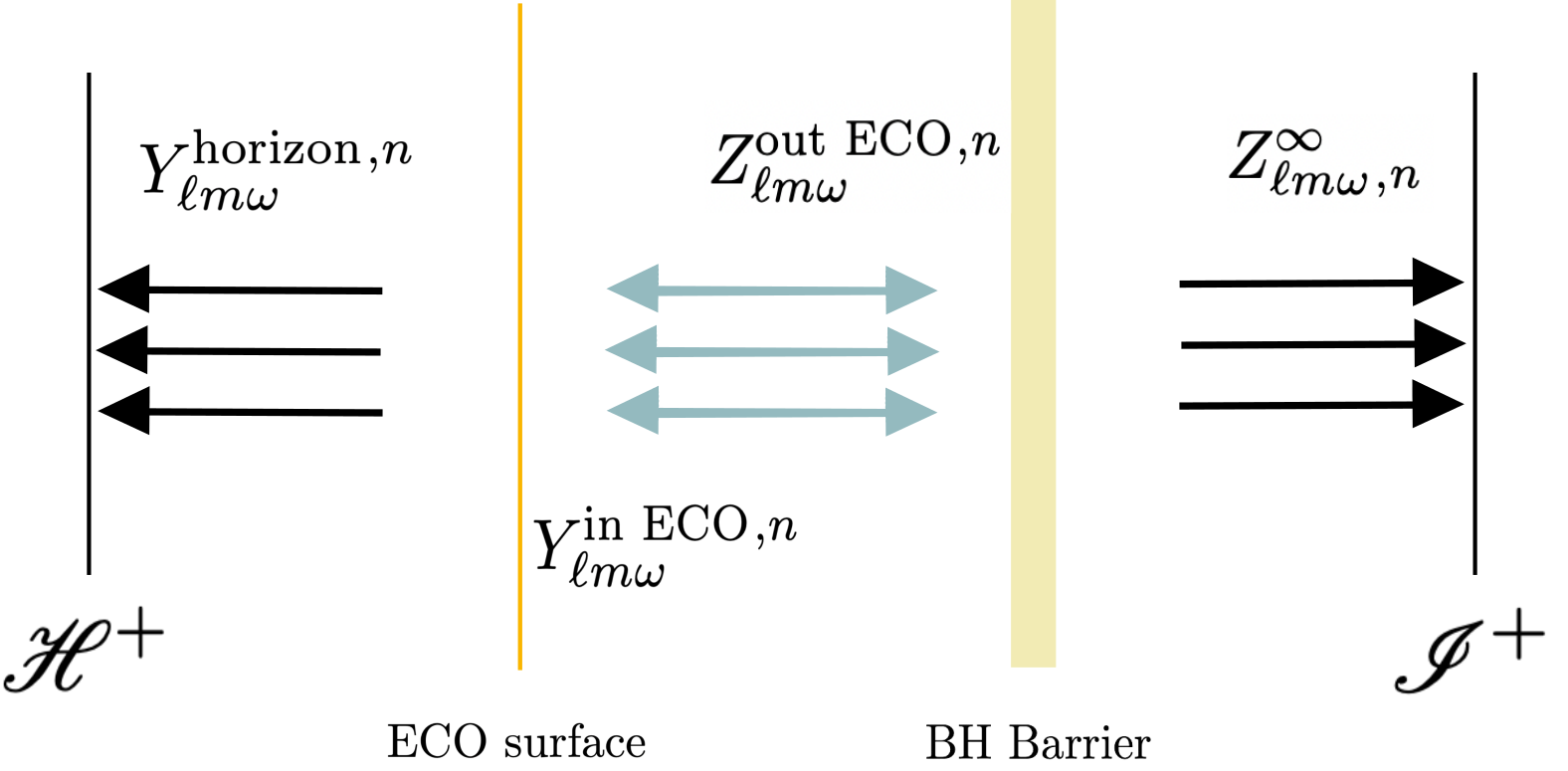}
	\caption{Sketch of the cavity formed between the future horizon and the BH barrier. The light blue arrows indicate the gravitational radiation trapped inside the cavity. The black arrows represent the GWs that escape and cross either of the future horizons, i.e., $\Hp$ or $\scrip$.}
	\label{fig:cavity}
\end{figure}

Considering the process of echo production in the cavity formed by the BH barrier and the ECO's (or QBH's) surface, it is natural to ask whether the involved GWs obey a form of (energy) flux conservation for gravitational radiation. This can be straightforwardly deduced by considering Fig. \ref{fig:cavity}. Staying consistent in our notation, we adopt the viewpoint of the ECO, but the same formalism holds for the QBH as well: The process starts with the ingoing gravitational radiation that first interacts with the reflective surface, i.e., $\Yineco{}^{,1}$. This radiation is partially reflected by the surface, producing the outgoing wave $\Zouteco{}^{,1}$, and partially transmitted as $\ZBH{}^{,1}$, which propagates towards the BH horizon $\Hp$. At the BH potential barrier, the outgoing wave $\Zouteco{}^{,1}$ can, again, either be reflected off or traverse the barrier. In the latter case, the radiation escapes to future null infinity, where it arrives as the first echo, $\Zinf{}^{,1}$, at the detector. The reflected portion, meanwhile, forms a new ingoing wave $\Yineco{}^{,2}$, initiating the second cavity cycle. This iterative mechanism underpins the production of successive echoes, with radiative information being continuously exchanged between the reflective surface and the BH potential barrier.
Propagating this configuration infinitely far into the future, one concludes that the energy flux carried by the initial strain directed toward the ECO's surface must equal the sum of the energy flux passing through both horizons, $\Hp$ and $\scrip$, i.e., 
\begin{align}
\label{equ:energy_balance}
    \frac{\dd E^\T{out ECO}}{\dd t}+ \frac{\dd E^\T{horizon}}{\dd t} = \frac{\dd E^\T{in ECO}}{\dd t} \,.
\end{align}
Naturally, the energy flux conservation equation can be expressed in terms of the amplitudes at each stage of the GW's interaction with the barriers or horizons. For instance, combining the configuration depicted in Fig. \ref{fig:cavity} with the radial solutions close to the ECO, one could also write
\begin{align}
\label{equ:some_equ_}
    \frac{\dd E^\T{in ECO}}{\dd t}- \frac{\dd E^\T{out ECO}}{\dd t} =\frac{\dd E^\infty{}^\T{\,in}}{\dd t}- \frac{\dd E^\infty{}^\T{\,out}}{\dd t}\,,
\end{align}
which is similar to the formulations found in \cite{Chen_2021,Teukolsky_1974}. Note that in equation \eqref{equ:some_equ_}, the \textit{up solution} was complemented by a set of ingoing waves at infinity, $\Yinf{}^\T{in}$. Then, $E^\infty{}^\T{\,in}$ is proportional to the absolute of $(\Yinf{}^\T{in})^2$, while $E^\infty{}^\T{\,out}\propto|\Zinf|^2$. At this point, it is crucial to emphasize that the notation for the radial solutions might vary across literature, especially when the Sasaki-Nakamura (SN) formalism \cite{Sasaki_2003} is used instead of the Teukolsky framework. For comprehensive treatments of echo computations using the SN formalism, we refer to works such as \cite{Teukolsky_1974,Hughes_2000,Xin_2021,Chen_2021}. 

For the fluxes reaching $\scrip$, the definitions given in equation \eqref{equ:fluxes} in combination with the asymptotic strain yield the correct result. In case the asymptotic strain is not directly available, we can use equation \eqref{equ:strain_to_Z} to extract it from $\Psi_4^\circ$. Defining the flux that falls down into the BH or ECO, or the initial flux directed towards the ECO's surface, on the other hand, is less trivial. Concretely, for the energy carried by $\Yineco$ one needs to convert $\Psi_4$ to $\Psi_0$ (as it is associated with the perturbations of the shear on the horizon) using the TS identities after which the energy flux can be derived from the area change according to Hartles formula of BH area increase. After a lengthy computation, one finds \cite{Teukolsky_1974, Chen_2021}
\begin{align}
\label{equ:energy_dE_in_ECO}
    \frac{\dd E^\T{in ECO}}{\dd \omega} = \sum_{\ell,m} \frac{\omega}{64\pi k (k^2+4\epsilon^2)(2r_+)^3}|\Yineco|^2\,,
\end{align}
where $k = \omega - m\frac{a}{2Mr_+}$,  $r_+ = M + \sqrt{M^2+a^2}$ and $\epsilon = \frac{\sqrt{M^2-a^2}}{4Mr_+}$. For the events relevant to this article, the dimensionless spin $a$ of the ECO/BH is (mostly) negligible, thus $k\approx \omega$ and $r_+ \approx 2M$. A similar result can be obtained for the reflected quantity $\Zouteco$ (see Fig. \ref{fig:cavity}), i.e.,
\begin{align}
    \frac{\dd E^\T{out ECO}}{\dd \omega} = \sum_{\ell,m} \frac{\omega}{4\pi k (k^2+4\epsilon^2)(2r_+)^3}|\Zouteco|^2\,.
\end{align}
With the defined reflectivities $\Reco$ and $\RBH$, the above energies can be related as follows:
\begin{align}
\label{equ:first_relevant}
    \frac{\dd E^\T{out ECO}}{\dd \omega} =|\Reco|^2\frac{\dd E^\T{in ECO}}{\dd \omega}\,,
\end{align}
and thus 
\begin{align}
\label{equ:some_stuff_echo}
    \frac{\dd E^\T{echo}}{\dd \omega} &= (1-|\RBH|^2)\frac{\dd E^\T{out ECO}}{\dd \omega}\notag \\&= (1-|\RBH|^2)|\Reco|^2 \frac{\dd E^\T{in ECO}}{\dd \omega}\,.
\end{align}
Finally, we can express the energy flux across $\Hp$ in terms of the same quantity at $\scrip$, namely,
\begin{align}
\label{equ:energy_law_the_best}
     \frac{\dd E^\T{horizon}}{\dd \omega} &= |\Teco|^2 \frac{\dd E^\T{in ECO}}{\dd \omega} \,,
\end{align}
where $E^\T{in ECO}$ can be replaced using equation \eqref{equ:some_stuff_echo}.
The energy loss per cycle in the cavity depicted in Fig. \ref{fig:cavity}, as a function of frequency, then corresponds to
\begin{align}
    \label{equ:energy_loss_per_cycle}
     &\Delta E^\T{loss} (\omega)=  \int\dd \omega \frac{\dd E^\T{horizon}}{\dd \omega} + \int \dd \omega \frac{\dd E^\T{echo}}{\dd \omega} \notag\\
    &=\int \dd \omega \left(\underbrace{|\Teco|^2}_{=:\Gamma} + \underbrace{|\Reco|^2 |T^\T{BH}|^2}_{=:\Theta} \right) \frac{\dd E^\T{in ECO}}{\dd \omega}\,,
\end{align}
where $|T^\T{BH}|^2:= 1 - |\RBH|^2$. The coefficients $\Gamma, \Theta$ represent the fraction of energy loss due to gravitational radiation crossing the ECO/QBH horizon and the potential barrier, respectively. Equation \eqref{equ:energy_loss_per_cycle} enables the explicit computation (and comparison) of the energy crossing the horizon $\Hp$ for a given event. The latter becomes particularly relevant in the context of ECOs as there may exist instances of those collapsing to form a BH after the first echo (first cycle) \cite{Chen_2019}. If the BH is considered in a classical framework, no echoes will be received by the detector at $\scrip$. Considering the formed BH to be represented by a QBH, the transfer function of subsequent echoes changes, leading to a distinct echo morphology. 

Analogous conservation laws can be established for the angular momentum flux towards the future horizons $\Hp$ and $\scrip$. Together, the energy and angular momentum fluxes across the BH horizon influence the entropy of the dynamical BH during the ringdown phase, thereby affecting its area in accordance with the first law of \gls{bh} thermodynamics. Therefore, in principle, it should be feasible to relate the energy and angular momentum flux induced by the ringdown to the entropy of dynamical BH (for recent progress see\cite{hollands2024entropy}). We will defer such an investigation to future work.



\section{Numerical Study}
\label{sec:numerics}
With the analytical descriptions of the GW echo and its associated memory in hand, we now aim to quantify its impact on the waveform detected by relevant GW interferometers. Specifically, we focus on \gls{lisa}, which is particularly well-suited for potential memory detections due to its high sensitivity in the low-frequency band, where the memory effect is typically stronger. To evaluate the \gls{snr} of the echoes and the corresponding memory effects, we utilize the pipeline developed in \cite{Henri_mem}. Thereby, a realistic prognosis for echo memory detection necessitates consideration of various event-related factors, including the orientation and sky position of the merger relative to the LISA frame. For the sky position and all other orientation-dependent features pertinent to the SNR, we adopt the conservative baseline parameters outlined in Table I of \cite{Henri_mem}.

In this section, we utilize multiple \textit{SXS} simulations with vanishing or nearly vanishing spin, that are \textit{SXS:BBH:0205}, \textit{0206}, \textit{0207}, \textit{1424}, \textit{1448}, \textit{1449}, \textit{1455}, \textit{1936}. Exceptions to the low-spin prescription are given \textit{SXS:BBH:0334}, \textit{1155}, and \textit{2108} which have a remnant spin amplitude $|\vec{\chi}|$ between $0.28$ and $0.68$. While we acknowledge that the algorithm computing the echoes may introduce systematic errors for these events, they will only play a secondary role in this analysis as corrections for non-trivial spin amplitudes are generally small. Regarding the reflectivity models, a non-negligible spin introduces line-broadening as discussed above and in \cite{Agullo_2021}. For the relevant cases however, the effective broadening remains close to the numerically imposed frequency resolution of the applied pipeline, i.e., $\mathcal{O}(1)$ $\mu$Hz. Thus, without loss of generality, we include the corresponding events into our consideration serving as consistency checks against potential parameter biases regarding the spin of the remnant object. For each event, the echo is numerically computed solely for the harmonic strain modes $h_{2,\pm2}$ due to their overall dominance. Note however that, particularly for the flux computation, all modes are included. Thus not all strain modes subject to the calculations below carry an echo. The implications of restricting the echo to the $h_{2,\pm2}$ mode will be addressed below.

\subsection{Echo-induced Gravitational Wave Memory}
\label{subsec:numerics_mem}
\begin{figure}
\includegraphics[width=1.0\columnwidth]{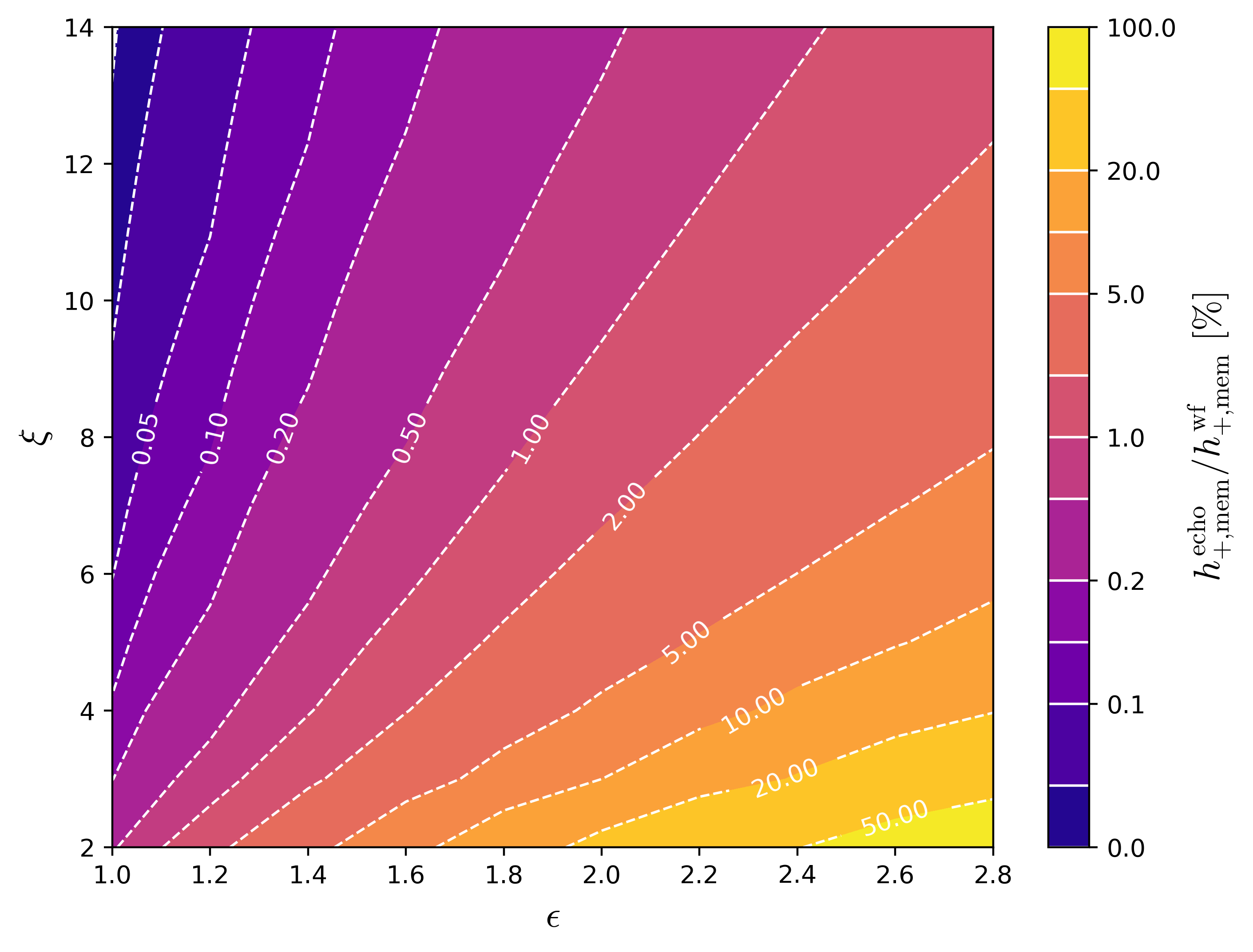} 
\includegraphics[width=1.0\columnwidth]{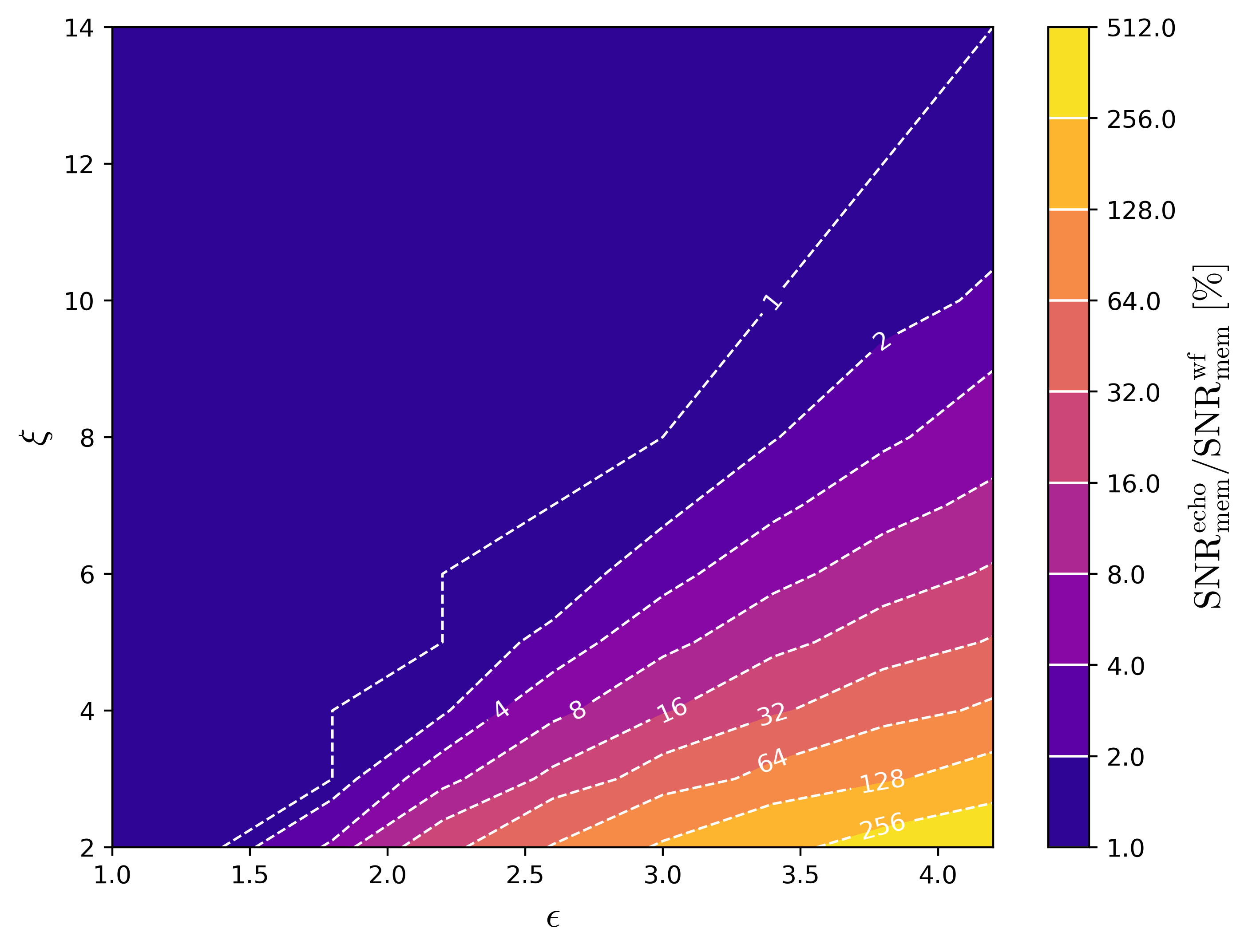}
     \caption{Memory amplitude and SNR gain for event \textit{SXS:BBH:1936} under the reflectivity model of a QBH with $\alpha=8\pi$, $\beta = 10^{-15}$ and $\delta=0.1$. The top plot shows the memory amplitude purely due to the echo as a fraction of the classical waveform memory (without echo). The bottom plot shows the gain in overall memory SNR relative to the SNR of the waveform memory without echo (at redshift $z=1$ and mass $M=10^6 M_\odot$ the total value for the SNR for the chosen baseline is roughly 10). Note that for better readability, the $x$-axis of the bottom plot is extended up to $\epsilon=4$.}
       \label{fig:numerics_I}
\end{figure}

To get an intuition for the significance of the echo-induced memory based on the QBH model, we simulate different reflectivity functions and compute the memory for each listed waveform. The results are exemplarily displayed for \textit{SXS:BBH:1936} in Fig. \ref{fig:numerics_I}. The event is chosen to match the results displayed in the companion paper \cite{Maibach_2024} on investigations regarding the general echo detectability prospects. 
The contour plots displayed in Fig. \ref{fig:numerics_I} show the fraction of memory attributed solely to the echo in comparison to the classical waveform's memory (top panel) and the corresponding \gls{snr} associated with the additional memory contribution (bottom panel)\footnote{Note the difference scales of the $\epsilon$-axis for better readability of the bottom planel.}. This \gls{snr} pertains exclusively to the detectability of the memory and is independent of the waveform itself. To facilitate a clearer interpretation, we present the relative increase in memory, thereby effectively factoring out the redshift and mass dependencies. For a comprehensive exploration of the detectability of memory concerning the latter two remnant-specific parameters, we refer to \cite{Henri_mem}. As anticipated, Fig. \ref{fig:numerics_I} reveals a discernible trend indicating that lower suppression factors, $\xi$, and $\epsilon$, are associated with higher echo memory contributions. Therefore, the echo's memory contribution is proportional to the its amplitude. Furthermore, the contour plot indicates that over a broad range of parameter space, the echo significantly contributes to the overall memory, achieving \gls{snr} levels comparable to the initial waveform's memory in extreme cases.
While the \gls{snr} for the echo memory alone is likely to be insufficient for individual detection, it is plausible that the echo memory enhances the overall memory \gls{snr}. This enhancement is heavily contingent upon the specific event and the temporal separation between the echo and the ringdown of the waveform. The sensitivity to time separation is primarily due to \gls{lisa}'s response with respect to the step-like increase characteristic to the memory. The closer the two memories, sourced by the waveform and the echo, reside in time, the more pronounced the features' low frequency content appears in \gls{lisa} data (see \cite{Henri_mem} for more details). Further, as elaborated above, terms such as $\dot{\bar{h}}^\infty\dot{h}^\T{echo} + c.c.$ now appear in the non-linear memory leading to a non-linear increase. Conversely, if this time separation is too pronounced, the synergistic effects between the waveform's and echo's memory diminish. This phenomenon is particularly evident in the analysis of the ECO model, which will be discussed further below.

It is crucial to highlight that, in parameter regions where the echo memory constitutes more than $\mathcal{O}(10)$ percent of the waveform memory, the oscillatory part of the echo's time series itself becomes large enough to fall within the detectable range of LISA. This detectability heavily relies on factors such as redshift and mass, which were explored in \cite{Maibach_2024}.  For a reference on the \textit{pure} echo SNR for a parameter region comparable to that displayed in Fig. \ref{fig:numerics_I}, we refer to in Fig. 5 therein. In this context, we find that for larger $\epsilon$ and smaller $\xi$'s, the echo's magnitude increases significantly faster than the memory contribution. One finds that once the echo amplitude reaches a threshold of approximately $\gtrsim 5\%$ of the waveform amplitude, \gls{lisa}'s \gls{snr} becomes sufficiently high to enable individual echo detection (given the right redshift and mass range of merger events), rendering the echo-induced memory as a primary tool of echo detection redundant. Nevertheless, we find it surprising that an event of such short time span compared to the original waveform exerts such large contributions to the memory.

\begin{figure}
\includegraphics[width=1.0\columnwidth]{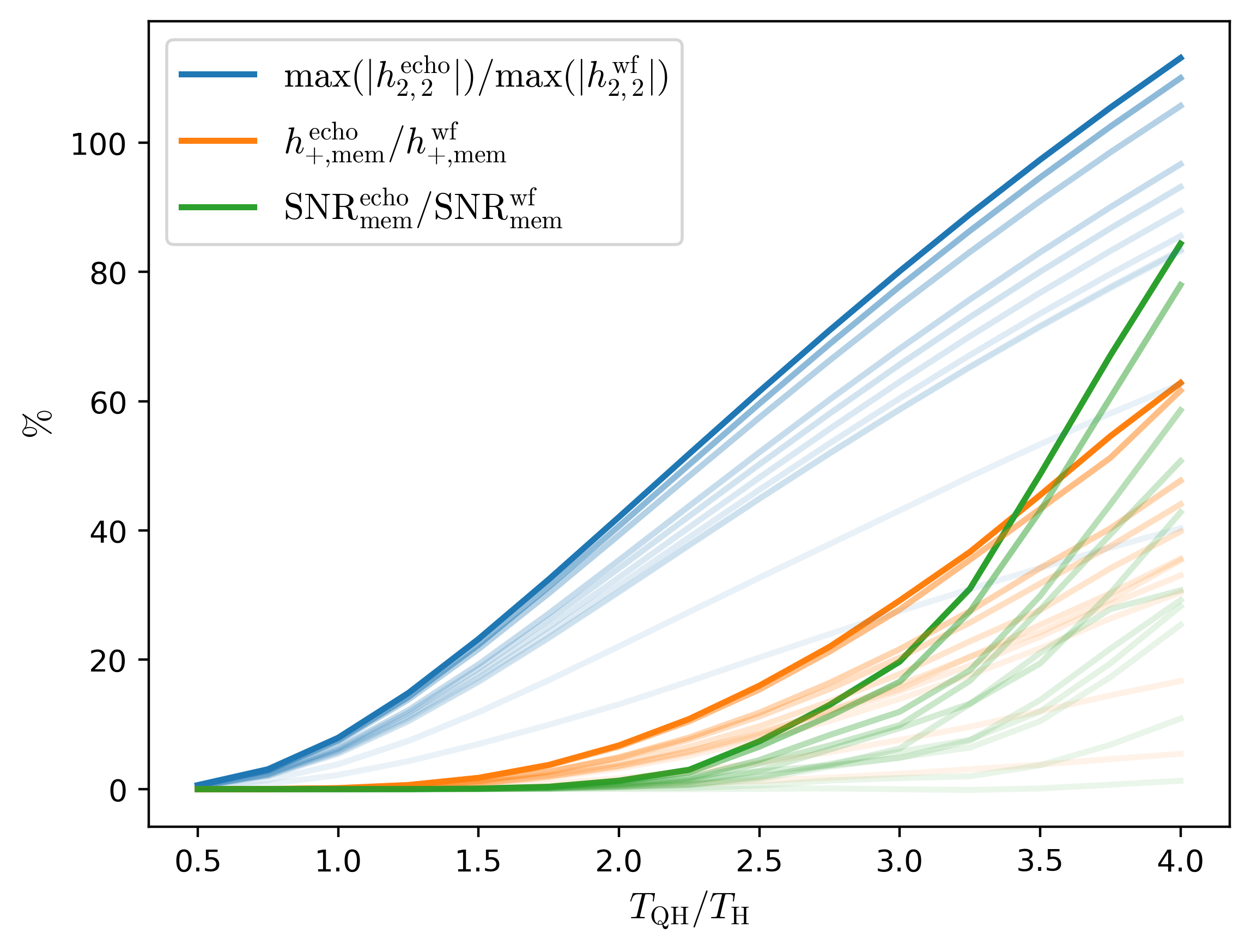}   
     \caption{Comparison of the echoes maximal magnitude, memory, and memory SNR with respect to corresponding quantities for the classical waveform (without memory) for the ECO model. Each line corresponds to a given event of the list above, where the color intensity is arranged such that the lines fade according their value for the relevant fraction. The graphs represent the ECO analog to Fig. \ref{fig:numerics_I}.}
       \label{fig:numerics_III}
\end{figure}

Transitioning to the ECO model, we present the parameter dependence of the echo memory contribution, its SNR, and the echo amplitude in Fig. \ref{fig:numerics_III}. The maximal echo amplitude here serves as a gauge for comparisons between echo memory and pure echo SNR. Due to its simplified parameter dependence, instead of focusing on a single event, we include all SXS simulations listed above in the plot. It is evident that while all events exhibit a trend similar to that of \textit{SXS:BBH:1936} for the QBH, there are notable event-dependent variations in the echo and echo memory amplitude ratios. These variations arise from distinct simulation parameters such as mass ratios and initial spins, leading to a diverse range of QNM content and, consequently, echoes. Fig. \ref{fig:numerics_III} further displays crucial difference between the reflectivity models and their associated parametrization. Specifically, while the response and transfer functions for the ECO and QBH are relatively similar (as illustrated in Fig. \ref{fig:Transfer_functs}), an increase in $T_\T{QH}$ for the ECO not only affects the exponential suppression but also compresses the time interval between the waveform ringdown and the echo. This results in a non-linear response of LISA in terms of the memory \gls{snr} in relation to the linear increase in the amplitude of the echo memory contribution when $T_\T{QH}$ is varied. A similar phenomenon can be observed for the QBH if $\beta$ is adjusted accordingly.

Most importantly, when comparing the results for the QBH model and the ECO, we find no significant differences in the memory contributions, provided that the parameters are selected to ensure comparable amplitudes of the transfer functions. This observation suggests that, despite the distinct underlying physics of the two models, their impact on gravitational wave memory is nearly identical. Validating this hypothesis further, we examine the remaining parameters, i.e., $\gamma$ in $\Reco$ as well as $\alpha,\beta$ and $\delta$ in $\RQBH$. While $\gamma$ and $\beta$ primarily influence the QNMs of the ECO and the QBH respectively (not to be confused with the QNMs of the ringdown, which feed the echo), $\alpha$ and $\delta$ modulate the position of the characteristic frequencies as well as the distinctiveness of their cusps. Our findings confirm that, as anticipated, none of these parameters significantly affect the echo memory with the exception of $\delta$. If large enough, $\delta$ can reduce the amplitude of the echo memory due to the increased absorption by the BH horizon. Although this feature would enable a distinction between reflectivity models solely based on the amplitude of the resulting echo memory, given the current detection estimates, it is unlikely that this amplitude can be determined with sufficient accuracy. 

For the tests of different reflectivity model parameters, the following parameter ranges were considered: $\gamma \in [10^{-15}, 10^{-4}]$, $\alpha\in [4 \log 2, 8\pi]$, and $\delta\in[0,1]$. The upper bound for $\gamma$ was chosen such that the separability condition remains intact, the lower bound is arbitrary as it would simply delay the echoes arrival further. For $\alpha$, we chose the interval depending on relevant values in literature \cite{Agullo_2021}. Finally, for $\delta$ the lower bound marks the level at which ECO and QBH transfer functions are indistinguishable at high frequencies. The upper bound establishes a critical value beyond which the characteristic frequencies' cusps are sufficiently pronounced to impact the transfer function's overall amplitude.

\subsection{Energy and Momentum Flux across $\scrip$}
\label{subsec:numerics_fluxes}
\begin{figure}
\includegraphics[width=1.0\columnwidth]{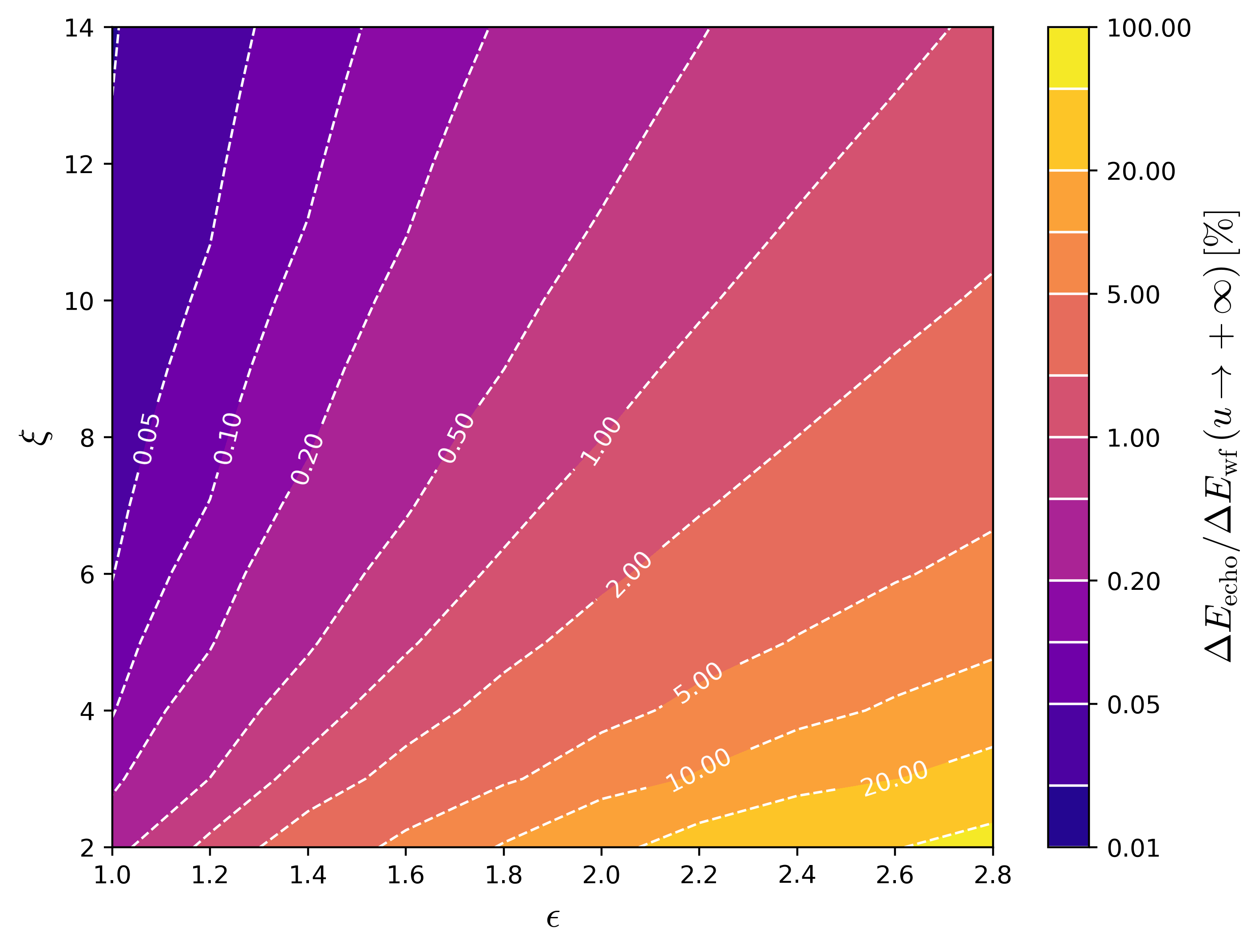} 
\includegraphics[width=1.0\columnwidth]{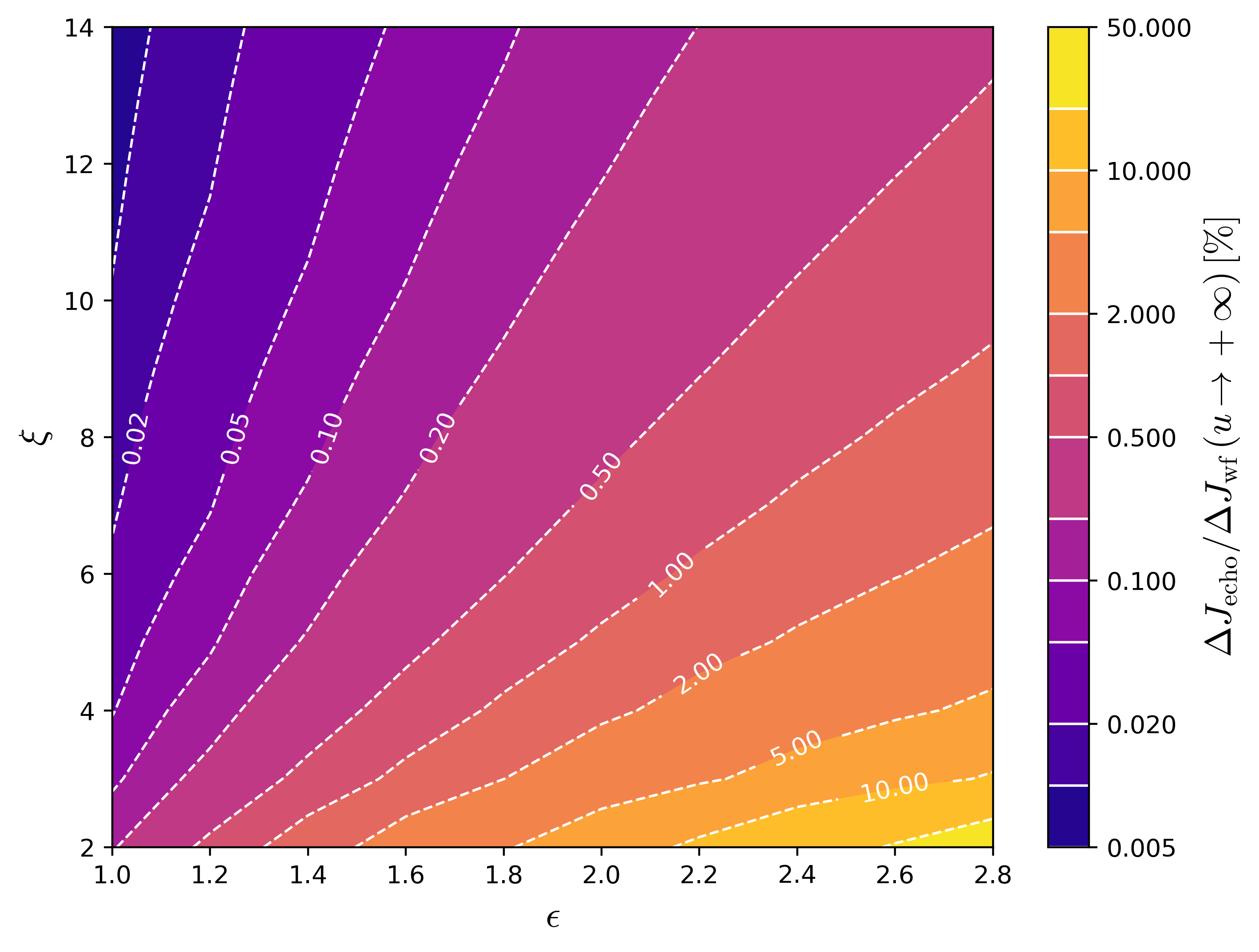}
     \caption{Integrated energy (top) and momentum (bottom) flux for \textit{SXS:BBH:1936} under the reflectivity model of a \gls{qbh}. For the parameters not displayed in the plots, the same values as in Fig. \ref{fig:numerics_I} have been chosen. }
       \label{fig:numerics_IV}
\end{figure}

To obtain a first characterization of the echo in terms of fluxes, we compare the time-integrated energy and momentum flux of the echo for \gls{eco} and \gls{qbh} against the corresponding fluxes of the waveform at $\scrip$. The relevant parameters are varied just as in the previous subsection. For the \gls{qbh}, Fig. \ref{fig:numerics_IV} displays the integrated energy and momentum fluxes measured \textit{morally} at ($u\rightarrow+\infty$). Numerically, this boundary is approximated by the maximum time extent of the simulated \textit{SXS} waveform.

\begin{figure}  
\includegraphics[width=1.0\columnwidth]{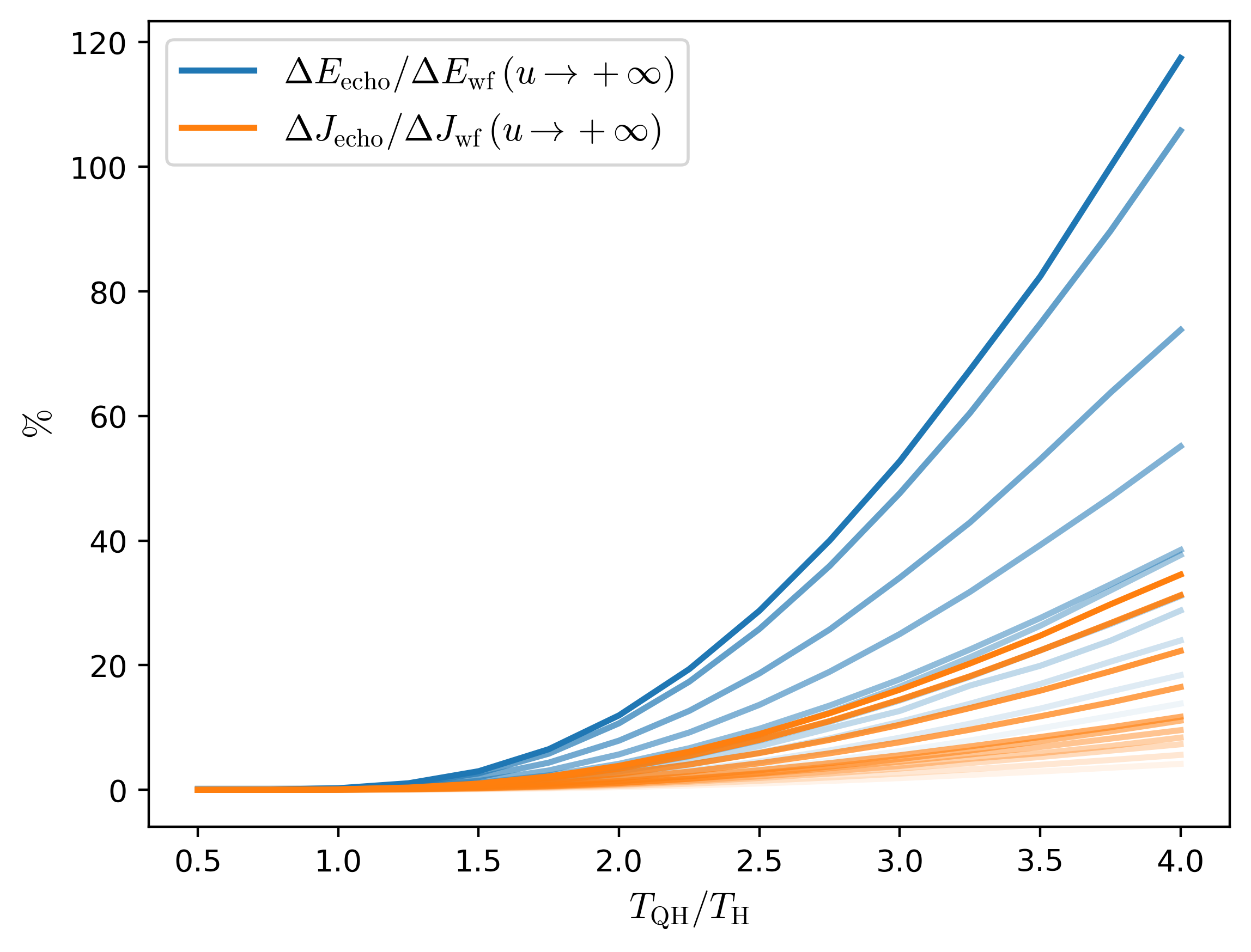}   
     \caption{Integrated energy and momentum flux for the listed events under the reflectivity model of an \gls{eco}. Again, each line corresponds to a given event of the list above, where the opacities are arranged such that the lines fade according to the largest value of the relevant fraction. The graphs represent the ECO analogue to Fig. \ref{fig:numerics_IV}.}
       \label{fig:numerics_V}
\end{figure}

The values presented in Fig. \ref{fig:numerics_IV} indicate that, within the parameter region for which the memory contributions is non-negligible, the fractional integrated energy flux exhibits behavior that closely resembles the fractional memory. Specifically, the energy flux associated with the echo can comprise a considerable fraction of the total energy carried by the complete waveform. A similar pattern is observed for the momentum flux; however, it is noteworthy that, for the tested events, the echo carries less momentum than energy. A similar relation between energy and momentum flux is evident for events with higher remnant spins as well, as demonstrated by Fig. \ref{fig:numerics_V}. Throughout these events, waveforms with larger energy flux also exhibit larger angular momentum flux, although the latter is generally much smaller in magnitude when compared to the full waveform's flux. We find that this behavior is uncorrelated with respect to the remnants spin. Physically, the observation that the echo carries less angular momentum flux than the initial waveform is sensible due to the generating process of the respective strain signals. While a coalescing binary is generally expected to radiate away large amounts of angular momentum, the perturbed Schwarzschild \gls{bh} is not.

In Fig. \ref{fig:numerics_V} the analysis of the \gls{eco} reflectivity model was extended to more extreme ratios $T_\T{QH}/T_\T{H}$ to demonstrate that the energy flux due to the echo can as well dominate the one of the waveforms. 
It's important to highlight that in this scenario, the individual echoes computed for the ECO model (see equation \eqref{equ:sum_echo}) transitions into a continuous signal due the decreasing time separation between individual echoes for larger $T_\T{QH}$. The strain associated with this situation is exemplarily illustrated in Fig. 12 of \cite{Ma_2022}. 

As for the memory, our findings indicate that the parameters $\gamma,\alpha,$ and $\beta$ do not substantially impact the energy or angular momentum flux when varied within the previously defined domains. Instead, Fig. \ref{fig:numerics_V} demonstrates that the fluxes carried by the echo strongly vary among simulated waveforms, leading to the conclusion that the direct and indirect features of \gls{gw} echoes are fundamentally determined by the classical QNM content of the remnant, given the reflectivity models presented here capture the physical reality of these systems sufficiently accurate. The latter is supported by the equations above, where $\Psi_{0, \ell m}(v)\mathcal{F}(v)$ represents the information encapsulated in the ingoing QNMs. This observations implies that, generally, the echo, its memory and associated fluxes is primarily determined by the physical properties of the remnant compact object. The reflectivity primarily affects the amplitude. Further support of this statement is provided by Appendix \ref{app:QNM}. Therefore, the selected models can be considered as robust in terms of potential errors due to additional phenomenology.

The above result equally holds when the echo is computed for all numerically accessible strain modes instead of only $h_{2,\pm2}$. Naturally, the fluxes increase slightly when the echo's mode content is extended. This, however, has to be anticipated given the sum over strain modes in the flux determining factors $\alpha^\T{echo}$ ($\beta^\T{echo}$) in equation \eqref{equ:alphas} (equations \eqref{equ:alphas_0} and \eqref{equ:alphas_1}).

Finally, we turn to the computation of an energy balance \eqref{equ:energy_loss_per_cycle} that can be established for the above fluxes. We are particularly interested in how the energy measured at the GW detector compares to the energy absorbed by the compact object or the BH at $\Hp$. 
To establish a flux balance, we compute equation \eqref{equ:energy_dE_in_ECO} and, in combination with equations \eqref{equ:first_relevant}-\eqref{equ:energy_loss_per_cycle}, estimate the energy consumed by the \gls{qbh}/\gls{eco} in comparison to the energy radiated to $\scrip$. Our findings reveal that, even when adjusting the reflectivity of the remnant object to boost the emission of large echoes, the overwhelming majority of energy - over $99\%$ of the total energy loss per cycle - is absorbed by the celestial body itself. In other words, for a stable cavity between the event horizon and the surrounding potential barrier, nearly all of the energy $E^\T{in ECO}$ is returned to the compact object, whether it is a \gls{qbh} or an \gls{eco}. 

To quantify the impact of characteristic absorption frequencies as the major distinguishing factor between the reflectivity models studied in this work, the reflectivity parameter $\delta$ is varied. Thereby, the remaining parameters are fixed such that the \gls{eco} and \gls{qbh} models yield identical transfer functions up to the cusps associated with $\omega_N$. It is found that the flux is highly sensitive to $\delta$. Already for $\delta=0.01$, the \gls{qbh}s flux decreases by $\mathcal{O}(1)$ percent with respect to the \gls{eco}s flux. This trend continues roughly linearly and holds equally for both energy as well as momentum flux. By increasing $\delta$ further, the \gls{qbh} reflectivity dramatically decreases, until, for $\delta\gg 1$, the remnant turns into a classical \gls{bh} as all energy is absorbed by $\Hp$. The sensitivity to $\delta$ suggests that the flux measurement or equally the amplitude of the echo-related memory contribution could in principle be used to distinguish the origin of echo-like features in the gravitational waveform, given the strain time series of such an echo reaches sufficient SNR. It is however unlikely that the precision on the echo memory amplitude outperforms the SNR of characteristic frequency features in LISA's \gls{tdi} data, as it is investigated in \cite{Maibach_2024}. A investigation of synergizing echo strain and memory in LISA data as well as detection prospects for more memory focused instruments is left to future work.

Finally, we test which of the cavity's ``walls'' acts as more confining regarding the energy traversing it. The latter can be quantified by the factors $\Gamma$ and $\Theta$ in equation \eqref{equ:energy_loss_per_cycle} denoting the energy contributions towards the BH horizon, $\Hp$, and future null infinity, $\scrip$, rewspectively. We find that the signal entering the cavity as $\Yineco$ in fact encounters the horizon of \gls{eco} and \gls{qbh} and the potential barrier with roughly equal integrated transmitivities\footnote{Here, as a reference, we chose the \gls{eco}s transitivity with $\gamma=10^{-15}$ and $T_\T{QH}=T_\T{H}$.}. The reflectivity, however, integrated over the frequency band is less than a $\mathcal{O}(1)$ percent, leaving the echo with a minimal amount of energy after escaping the cavity towards $\scrip$, as depicted in Fig. \ref{fig:cavity}. The large amount of absorbed energy implies for a large class of events, according to \cite{Chen_2019}, the \gls{eco} may transform into a \gls{qbh} within the first cycle of radiation traversing the cavity. This implies that even when features of the characteristic frequencies are detected either in form of cusps in \gls{tdi} data \cite{Maibach_2024} or in reduced echo memories, it is not immediately clear if the remnant object initially was a \gls{qbh} or a \gls{eco}. Therefore, the detection of echos with characteristic frequencies cannot not rule out the existence of \gls{eco}s.



\section{Discussion}
\label{sec:conclusion}
The \gls{gw} memory effect will likely enter the regime of detectability of future space-based detectors such as LISA \cite{LISA,Henri_mem} as well as of next-generation ground-based detectors such as Cosmic Explorer \cite{Cosmic_Exp} or the Einstein Telescope \cite{Einstein_Tel}. Besides testing the very fundamental nonlinearity of \gls{gr}, it simultaneously establishes a smoking gun for deviations beyond \gls{gr} \cite{Heisenberg_2023} and other subtle effects that can affect the permanent displacement of freely floating test masses. The primary challenge in detecting new and subtle features in gravitational waveforms lies in the need to generate highly precise \gls{gw} templates for extracting signals from noise-plagued data. Each additional feature incorporated into the template significantly expands the parameter space of the simulated waveform, resulting in a steep increase in computational costs. 
In this work, we explored one type feature induced by quantum phenomenology and the potential existence of \gls{eco}s that can easily be added to the waveform without drastic changes on waveform templates. The quantum phenomenological approach was rooted in basic area quantization arguments, which facilitated the development of a reflectivity model for \gls{qbh}s. This model was compared against established models for ECOs. For both models, corresponding GW echoes were constructed using the methodology outlined in \cite{Ma_2022}. We subsequently computed the memory and flux contributions of these echoes. We provided semi-analytical expressions for the gravitational memory effect induced by the echo as well as the corresponding fluxes. We further formulated an energy balance for the ingoing radiation towards $\Hp$ during and after the ringdown phase of the merger. Numerical investigation of both the echo-induced memory, focusing on its amplitude and \gls{snr}, and the flux balance were conducted, considering in-depth different scenarios within a reasonable parameter space and the resulting phenomenological consequences. 

Our results most notably reveal that memory corrections induced by GW echoes can reach levels of significant contributions with respect to the main waveform's memory. Although, for a large region of parameter space, the echo memory is too small to be detectable individually, it can however boost the total memories \gls{snr} considerably. Further synergies are observed when the time separation between the echo and the merger waveform are small. Physically, this corresponds to the reflective shell being located closer to the BH potential barrier. Thus, deviation from the expected memory could imply the existence of echos, offering an indirect detection avenue that does not require extensive waveform model adjustments. Given the results of \cite{Maibach_2024}, in practice a combined search of memory- and strain time series-related features of echoes is likely to provide the most robust detection prospects, given the dominance of time series SNR in comparison to the induced memory SNR for a given echo.
Generally, the time separation of echoes is highly model-dependent, and there is no consensus on its exact magnitude. Some studies even propose searching for \textit{rogue} echoes, which cannot be clearly associated with a specific merger event \cite{zimmerman2023rogueechoesexoticcompact}. Similar searches testing mainly outlier events or instrumental glitches have been proposed in other contexts as well, such as, for instance, in investigations of topological Dark Matter (see for instance \cite{Heisenberg:2023urf} and references therein).

Regarding the flux analysis, our investigation demonstrate that the majority of energy and angular momentum stored in the ingoing $\Psi_0$ heading towards $\Hp$ will ultimately be absorbed by the BH or \gls{eco}. Thus, one does not expect drastic phenomenological or astrophysical consequences from the reflective shells installed in this work besides the subtle echos in the interferometer data. Note further that the frequency spectrum considered in this work applies only to \gls{gw}s. For other types of radiation, the implications of, for instance, the area quantization arguments above may be dissimilar, see \cite{Agullo_2021} and references therein. Yet, despite transporting only a negligible fraction of energy and momentum flux of the ingoing $\Psi_0$ to future null infinity $\scrip$, the echo's contribution to the total energy and angular momentum of the waveform can still be significant. We further note that the echo consistently carries less angular momentum than energy compared to the classical waveform across all tested events. This trend is largely independent of the specific reflectivity parameters of the system and matches the expectations based on the particular dynamics that give rise to the echo. 

It is important to acknowledge that the echo memory's manifestation within the gravitational waveform is roughly model-independent. Solely its amplitude can differ based on the selection of the reflectivity model. In this work, the distinction between models was encapsulated in the reflectivity parameter $\delta$. Therefore, in principle, the echo-induced memory corrections establish an easy-to-add but powerful feature in waveform templates for future \gls{gw} detections. In this context, our findings suggest that the reflectivity models only play a secondary role for the exact shape of the echo time series as well as the associated memory. It is rather the dynamics of the binary configuration and the resulting QNM spectrum of the remnant body that determines the fluxes and $h^\text{echo}$. Measuring an echo therefore, in principle, establishes another pathway towards studying QNMs in the context of quantum \gls{bh}s.

The model-independence of the memory does not come without a price. We emphasizes at this point that a major challenge of identifying the echo-induced memory is its distinction with respect to other memory corrections. Such include significant corrections due to deviations from \gls{gr} \cite{Heisenberg_2023} and/or quantum effects unrelated to echoes \cite{Guerreiro_2022,Parikh_2021}. Further theoretical and analytical investigations of such effects would help identify distinct fingerprints of the potential features in both the memory and the oscillating part of the strain time series.


Another theoretical aspect that has been only partially addressed in this work, as well as in the companion paper \cite{Maibach_2024}, and that warrants further exploration, is the BH reflectivity model itself. Although the arguments laid out in Section \ref{sec:theoretical_background} (primarily based on \cite{Maibach_2024}) are consistent with phenomenological and observational constraints, advancing the study of gravitational wave GW echoes demands reflectivity models rooted in quantum information theory as well as astrophysics. Specifically, it is crucial to ensure that these models do not violate the core principles of BH information theory. In particular, the boundary conditions for \gls{qbh}s should be the subject of further investigation. One has to acknowledge that the prescription given in Section \ref{subsec:echo_reconstruction} is oversimplified. The ingoing radiation toward the BH horizon might, in fact, introduce a feedback term in the Teukolsky equation. Another alternative could involve GW scattering, rather than reflection off the horizon. These crucial aspects of QBH models must be thoroughly explored in future research to ensure a more accurate understanding of their physical behavior and the corresponding GW signatures.


In conclusion, we find that echoes and their signatures provide a smoking gun for quantum corrections to the \gls{bh}'s horizon as well as the existence of \gls{eco}s, provided a sufficient measurement precision. As their features appear in the non-linear GW memory, potentially among numerous other phenomenological imprints, the memory's detection has the potential to be the next milestone in gravitational physics and to fundamentally change our understanding of \gls{bh}s and other compact stellar objects.



\section*{Acknowledgements}
The authors wish to thank  Henri Inchausp\'e and Do\u{g}a Veske for productive consultations as well as the LISA Simulation Working Group and the LISA Simulation Expert Group for the lively discussions on all simulation-related activities.
LH would like to acknowledge financial support from the European Research Council (ERC) under the European Unions Horizon 2020 research and innovation programme grant agreement No 801781. LH further acknowledges support from the Deutsche Forschungsgemeinschaft (DFG, German Research Foundation) under Germany's Excellence Strategy EXC 2181/1 - 390900948 (the Heidelberg STRUCTURES Excellence Cluster). The authors thank the Heidelberg STRUCTURES Excellence Cluster for financial support and acknowledge support by the state of Baden-W\"urttemberg, Germany, through bwHPC. Research at Perimeter Institute is supported in part by the Government of Canada through the Department of Innovation, Science and Economic Development and by the Province of Ontario through the Ministry of Colleges and Universities. This material is based upon work supported by the National Science Foundation under Grants No. PHY-2407742, No. PHY- 2207342, and No. OAC-2209655 at Cornell. Any opinions, findings, and conclusions or recommendations expressed in this material are those of the author(s) and do not necessarily reflect the views of the National Science Foundation. This work was supported by the Sherman Fairchild Foundation at Cornell. This work was supported in part by the Sherman Fairchild Foundation and by NSF Grants No. PHY-2309211, No. PHY-2309231, and No. OAC-2209656 at Caltech.

\clearpage

\appendix

\section{Decomposition of Flux and Memory}
Energy and angular momentum flux as stated in equations \eqref{equ:fluxes} are both angle-dependent and thus possess a non-trivial decomposition in the bases of spin-weighted spherical harmonics. This choice of basis has been proven beneficial in some applications regarding gravitational waveforms \cite{DAmbrosio_2024} and, in general, allows for a more compact denotation. In this appendix, we provide an explicit decomposition of the terms appearing in the flux formulas \eqref{equ:fluxes}. We start by considering the energy, which contains contributions such as
\label{app:decomp}
\begin{align}
     |\dot h|^2 = \sum_{\ell, m} \alpha_{\ell m} Y_{\ell m}(\theta, \phi)\,,
\end{align}
which has spin-weight zero. It follows that 
\begin{widetext}
\onecolumngrid
    \begin{align}\label{equ:alphas}
        \alpha_{\ell m} = \sum_{\ell_1=2}^\infty \sum_{\ell_2=2}^\infty \sum_{|m_1|\leq \ell_1}\sum_{|m_2|\leq \ell_2} (-1)^{m_2+m} \dot h_{\ell_1 m_1} \dot{\bar{h}}_{\ell_2m_2}\sqrt{\frac{(2\ell_1+1)(2\ell_2+1)(2\ell+1)}{4\pi }}\begin{pmatrix}
\ell_1 & \ell_2 & \ell\\
m_1 & -m_2 & -m
\end{pmatrix}\begin{pmatrix}
\ell_1 & \ell_2 & \ell\\
2 & -2 & 0
\end{pmatrix}.
    \end{align}
\end{widetext}
For the decomposition of the angular momentum flux, we focus on two types of mixings where we neglect the time derivative as it does affect the spin-weighted basis. First,  we have
\begin{align}
    \bar h \eth h = \sum_{\ell, m} \beta_{\ell, m} \,_{1}Y_{\ell m}(\theta, \phi)
\end{align}
which is of spin-weight one and where we find 
\begin{widetext}
\onecolumngrid
    \begin{align}\label{equ:alphas_0}
        \beta_{\ell m} = \sum_{\ell_1=2}^\infty \sum_{\ell_2=2}^\infty \sum_{|m_1|\leq \ell_1}\sum_{|m_2|\leq \ell_2} (-1)^{m_1+m-1} \bar h_{\ell_1 m_1} h_{\ell_2m_2}\sqrt{\frac{(2\ell_1+1)(2\ell_2+1)(2\ell+1)}{4\pi(\ell_2+2)^{-1}(\ell_2-1)^{-1}}}\begin{pmatrix}
\ell_1 & \ell_2 & \ell\\
-m_1 & m_2 & -m
\end{pmatrix}\begin{pmatrix}
\ell_1 & \ell_2 & \ell\\
-2 & 1 & 1
\end{pmatrix},
    \end{align}
\end{widetext}
and similarly 
\begin{align}
     h \eth \bar h = \sum_{\ell, m} \Tilde{\beta}_{\ell, m} \,_{1}Y_{\ell m}(\theta, \phi)
\end{align}
with spin weight one as well and 
\begin{widetext}
\onecolumngrid
    \begin{align}\label{equ:alphas_1}
        \Tilde{\beta}_{\ell m} = \sum_{\ell_1=2}^\infty \sum_{\ell_2=2}^\infty \sum_{|m_1|\leq \ell_1}\sum_{|m_2|\leq \ell_2} (-1)^{m_2+m-1}  h_{\ell_1 m_1} \bar h_{\ell_2m_2}\sqrt{\frac{(2\ell_1+1)(2\ell_2+1)(2\ell+1)}{4\pi(\ell_2-2)^{-1}(\ell_2+3)^{-1}}}\begin{pmatrix}
\ell_1 & \ell_2 & \ell\\
m_1 & -m_2 & -m
\end{pmatrix}\begin{pmatrix}
\ell_1 & \ell_2 & \ell\\
2 & -3 & 1
\end{pmatrix}.
    \end{align}
\end{widetext}
The latter two coefficients cover all terms within the angular momentum flux \eqref{equ:fluxes}, as the four terms present in their differ only by a time derivative of $h_{\ell_1 m_1}$ or $h_{\ell_2 m_2}$.

\section{Parameter dependence of QBH echoes}
\label{app:A}
\begin{center}
\begin{figure*}[h!]
\subfloat[\label{fig:echo_a}]{%
  \includegraphics[width=1\columnwidth]{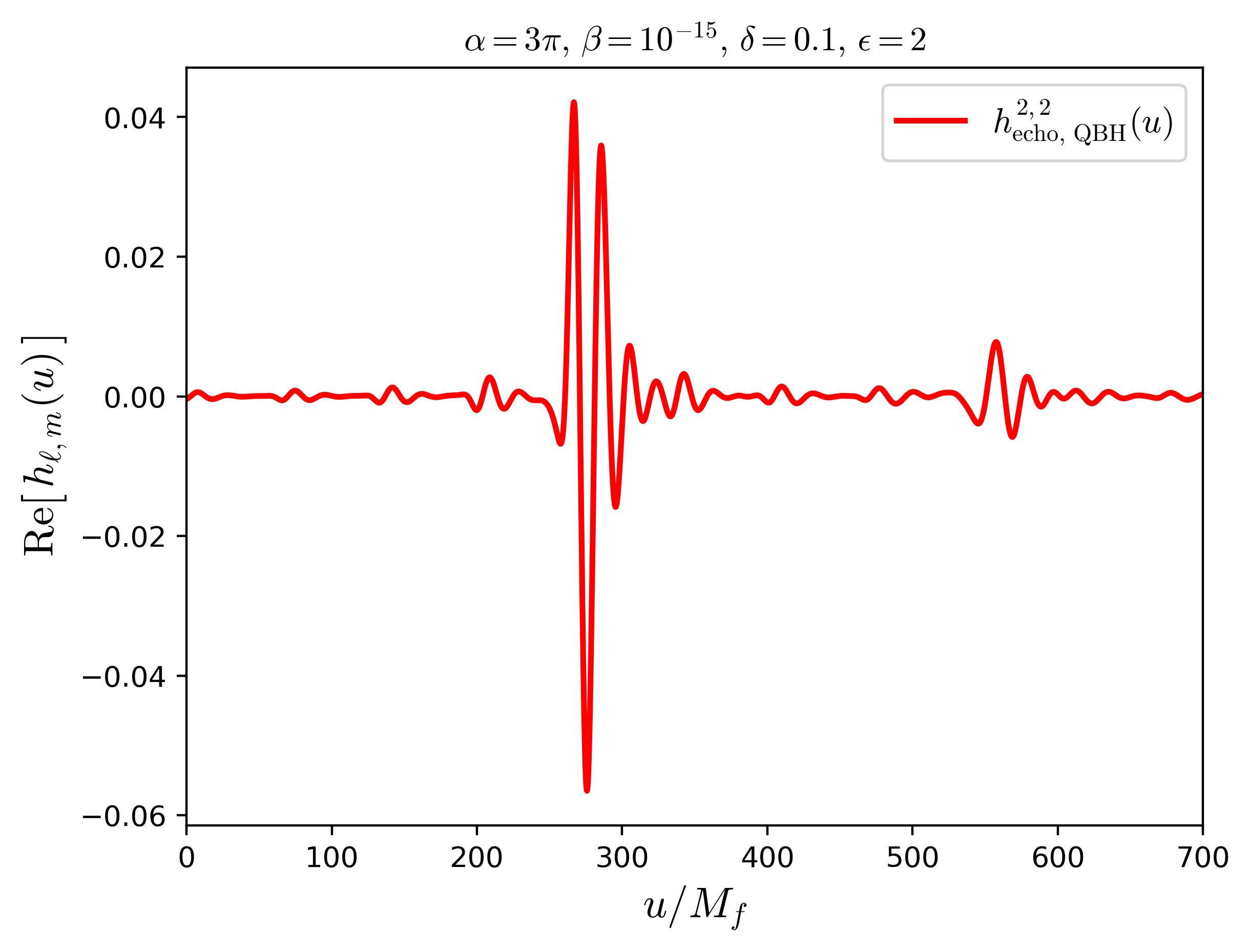}%
}\hspace*{\fill}%
\subfloat[\label{fig:echo_b}]{%
  \includegraphics[width=1\columnwidth]{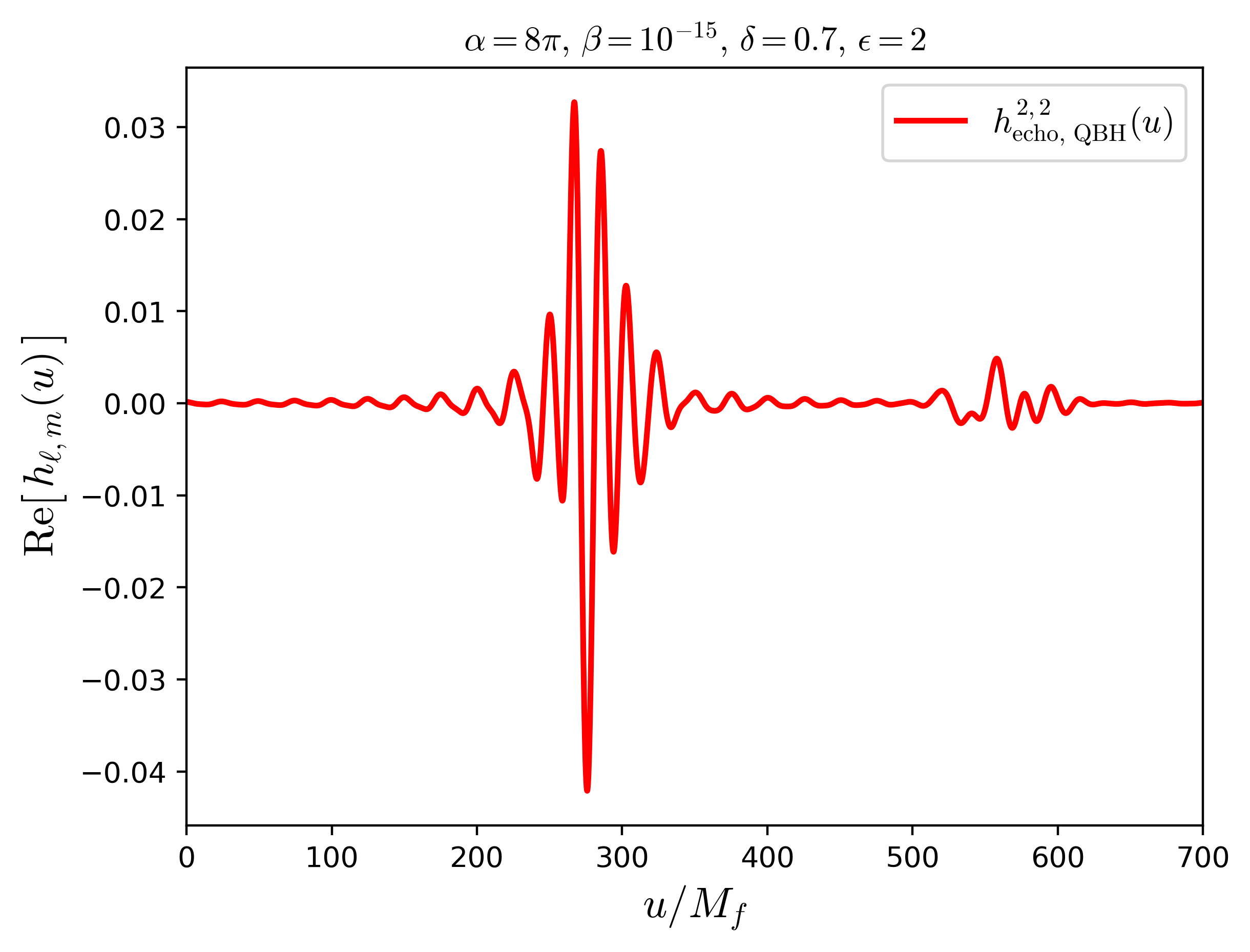}%
}

\medskip

\subfloat[\label{fig:echo_c}]{%
  \includegraphics[width=1\columnwidth]{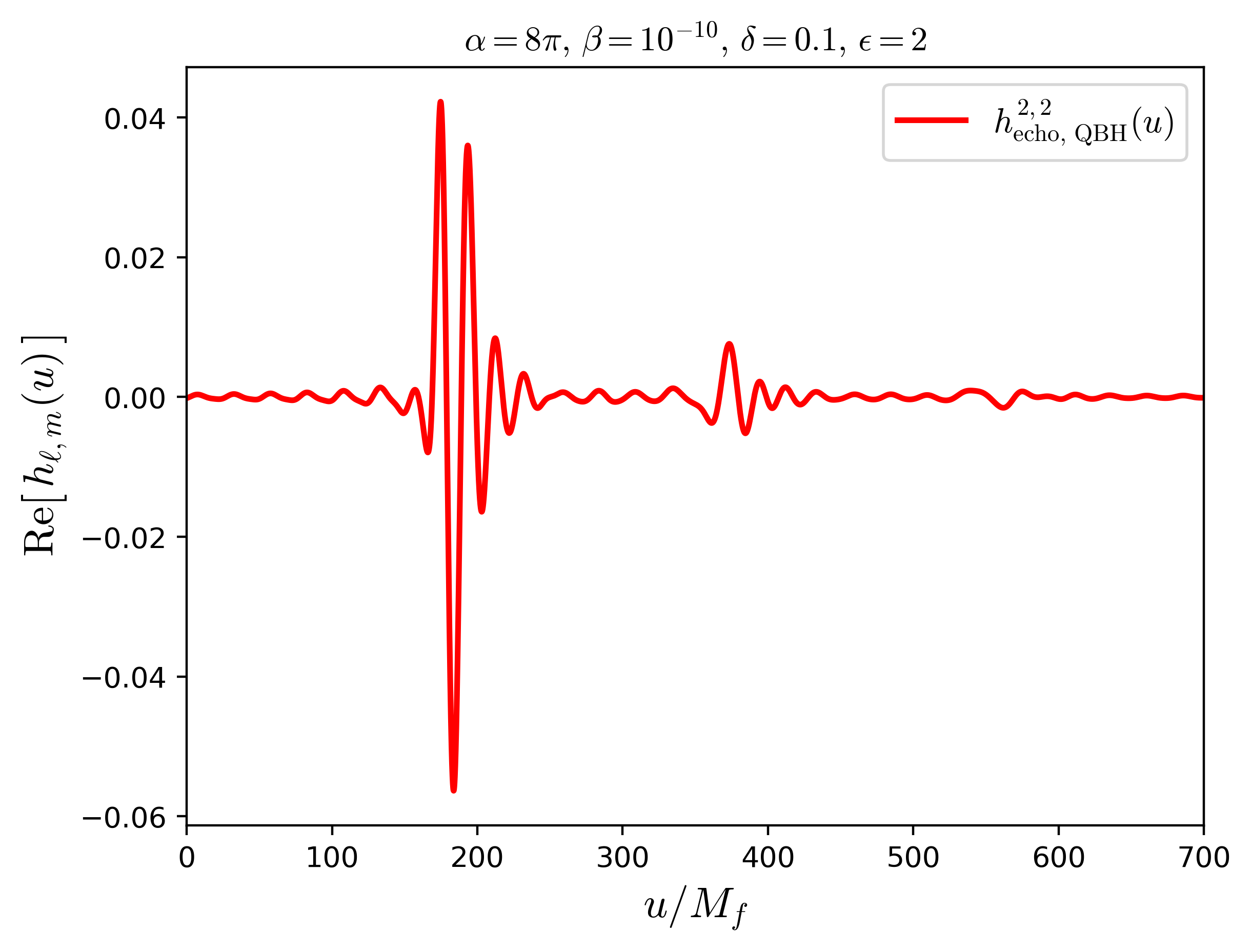}%
}\hspace*{\fill}%
\subfloat[\label{fig:echo_d}]{%
  \includegraphics[width=1\columnwidth]{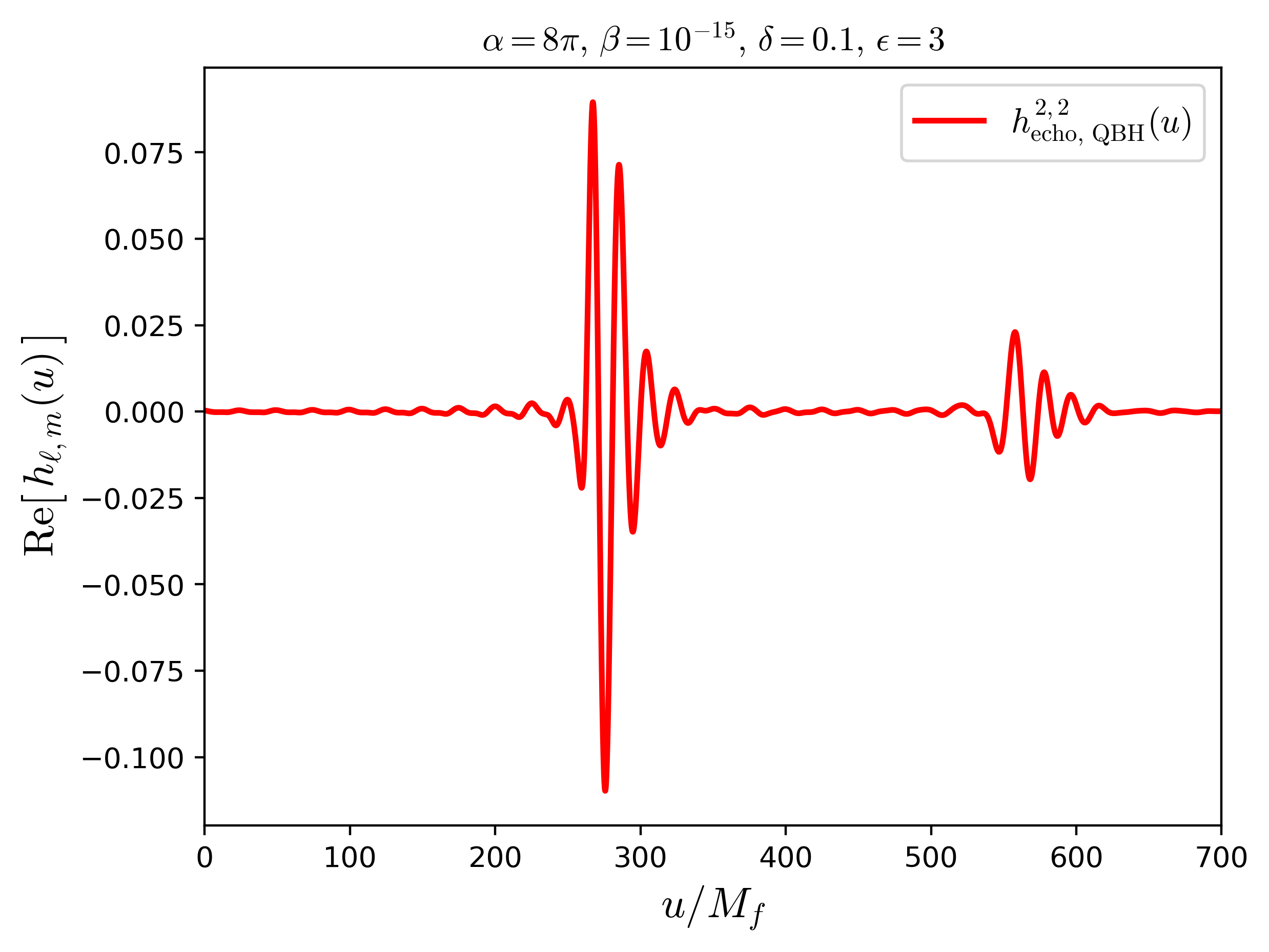}%
}
    
\caption{Pure echo strain $h_{2,2}^\T{echo}$ for the QBH computed using different model parameter and \textit{SXS:BBH:0207}.} \label{fig:echo_QBH}
\end{figure*}
\end{center}

The QBH echo is exemplarily shown for different parameter configurations in Fig. \ref{fig:echo_QBH}. The small noise-like oscillations are due to numerical issues with handling the transfer functions poles, representing the QNMs of the QBH. For the results of section \ref{sec:numerics}, these osciallations are irrelevant.

\section{Quasi-normal modes and overtones}
\label{app:QNM}
\begin{figure}  
\includegraphics[width=1.0\columnwidth]{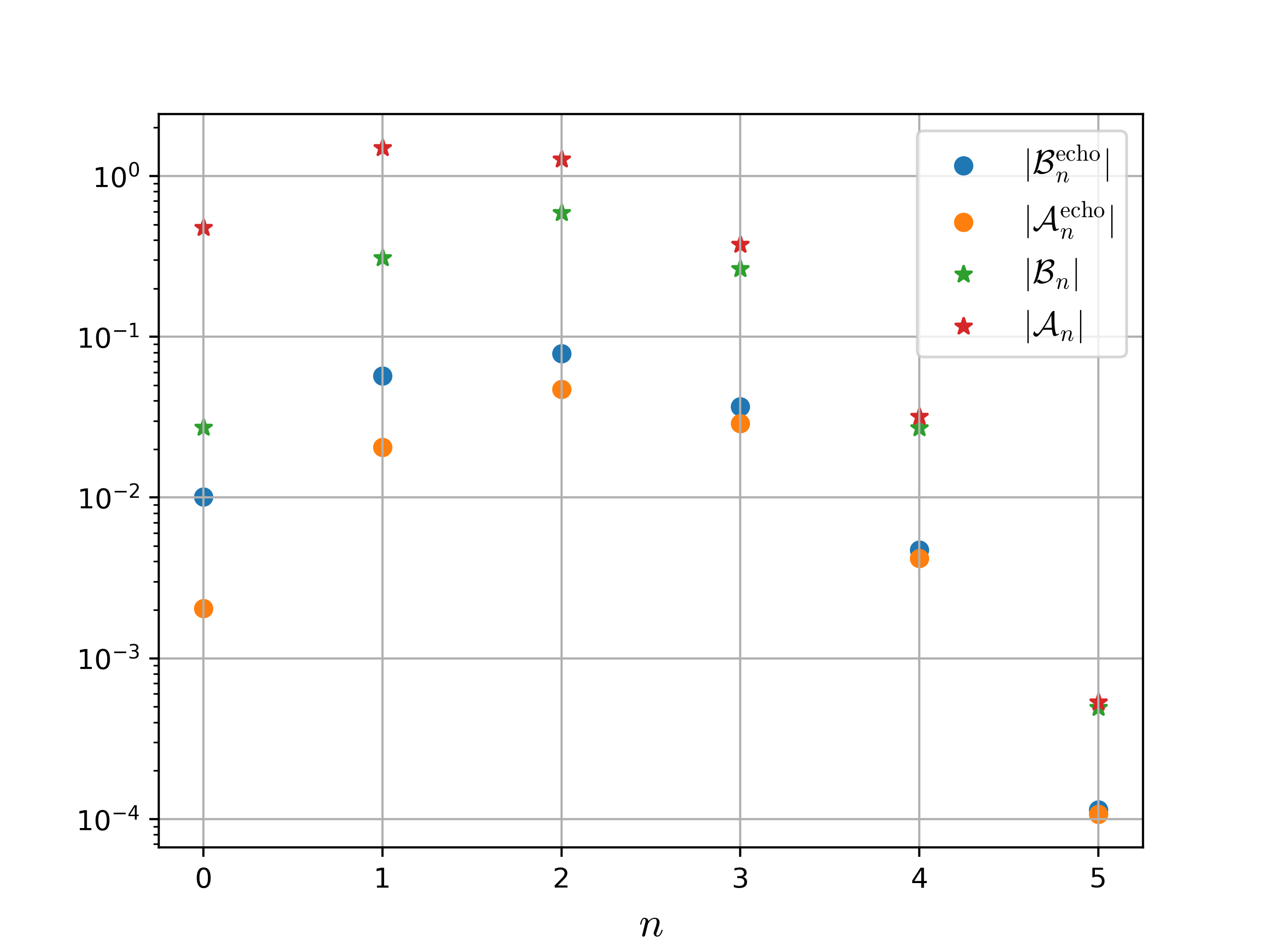}   
\includegraphics[width=1.0\columnwidth]{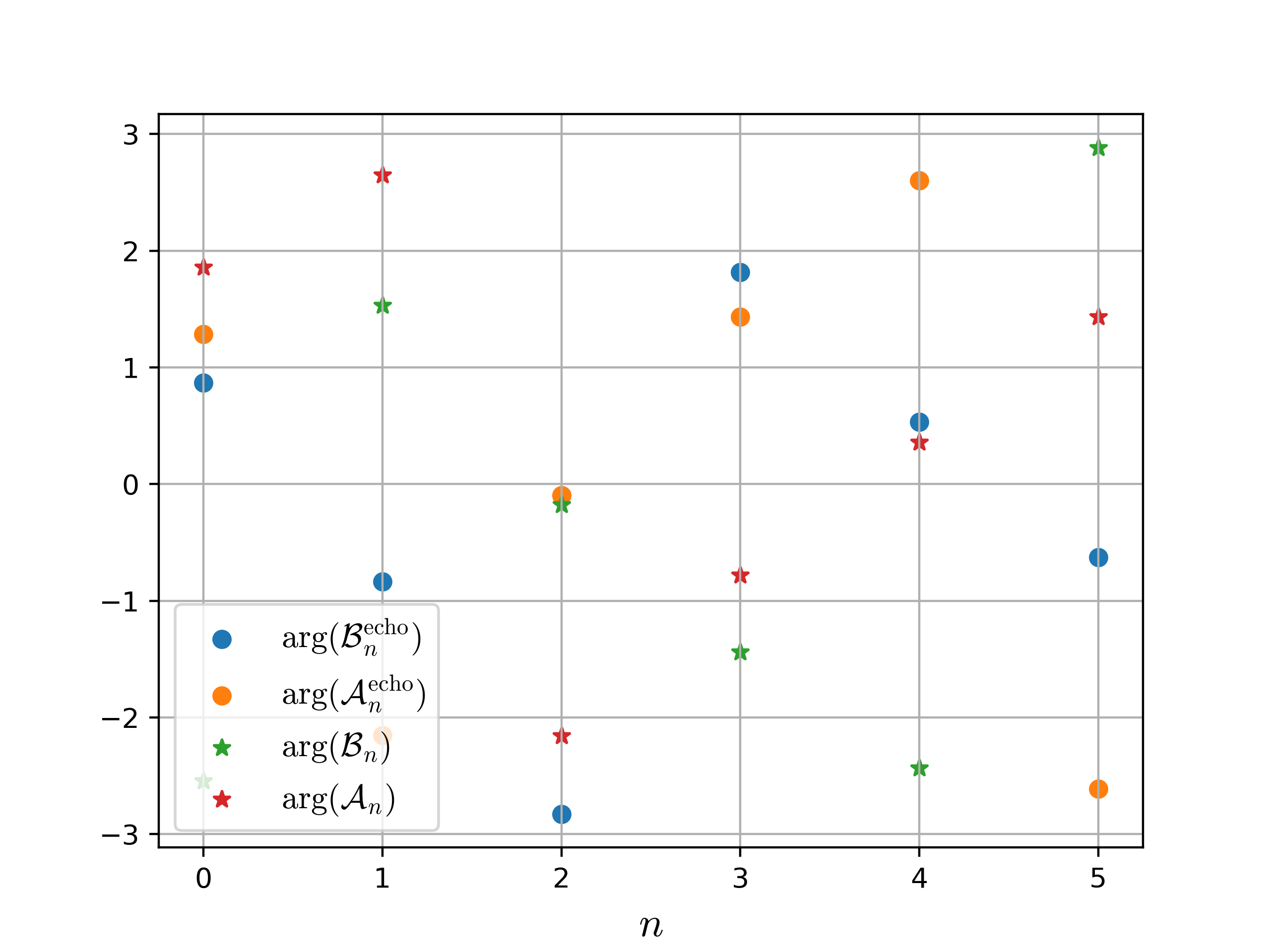}   
     \caption{Fitted overtones for $\Yin$ of event \textit{SXS:BBH:0207} and the echo resulting from the ECO reflectivity model with $T_\text{QH}=T_\text{H}$ and $\gamma = 10^{-15}$. The top plot shows the magnitudes of the overtones, the bottom plot the phases.}
       \label{fig:overtones}
\end{figure}

As described in the main text, the radiation that initially enters the cavity, $\Yin$ can be rewritten in terms of Schwarzchild QNMs $\omega_n$ and the corresponding overtones. One way of expanding $\Yin$ in the relevant time domain, i.e., for $v>v_\Sigma^H$ is given by
\begin{align}
\label{equ:overtones}
    Y^\text{in}_{22}(v>v_\Sigma^H) = \sum_{n=0}^{n_\text{max}}\left(\mathcal{A}_ne^{-i\omega_nv}+\mathcal{B}_ne^{i\overline\omega_nv}\right)\,,
\end{align}
where $\omega_n$ as well as the overtones $\mathcal{A}_n,\mathcal{B}_n$ are complex. The latter equation is used to determine $v_\Sigma^H$ in the echo construction pipeline via a least mismatch scheme. Using equation \eqref{equ:overtones}, the overtones can directly be related to the echo strain as 
\begin{align}
\label{equ:echo_overtone}
    h_{22\omega}^\text{echo}&=\frac{1}{\omega^2} \mathcal{K}(\omega) \mathfrak F\left[\sum_{n=0}^{n_\text{max}}\left(\mathcal{A}_ne^{-i\omega_nv}+\mathcal{B}_ne^{i\overline\omega_nv}\right)\mathcal(v)\right]\notag\\
    &= \frac{1}{\omega^2} \mathcal{K}(\omega)\sum_{n=0}^{n_\text{max}}\left(\frac{\mathcal{A}_n}{i(\omega+ \omega_n)}+\frac{\mathcal{B}_n}{i(\omega-\overline{\omega}_n)}\right)\,.
\end{align}
This implies that the amplitude and phase of the overtones is non-trivially modulated by the factor $\omega^{-2}\mathcal{K}(\omega)$. Since in this particular work, $\mathcal{K}(\omega)$ is not analytical, as exact analytical mapping between the modes and overtones of $h^\text{echo}_{22}$ and $Y^\text{in}_{22}$ is postponed to future explorations. Numerically, the latter is worked out in Fig. \ref{equ:overtones} by extracting the resulting echo modes via a similar fit analysis as for $\Yin$ in the main text. In practice, this corresponds to casting the echo time series into a form such as
\begin{align}
    h^\text{echo}_{22}(u>u_\text{echo}) = \sum_{n=0}^{n_\text{max}}\left(\mathcal{A}^\text{echo}_ne^{-i\omega_nu}+\mathcal{B}^\text{echo}_ne^{i\overline\omega_nu}\right)\,,
\end{align}
where $u_\text{echo}$ marks time after the merger at which the echo arrives at the detector. Note that, when applying this fit function the residual error in the mismatch is comparably large, i.e., of order $\mathcal{O}(0.1)$ percent. This is partially due to including only overtones up to $n_\text{max}<6$ in the fitting for both $Y^\text{in}_{22}$ an $h^\text{echo}_{22}$. Therefore, the values for the echo overtones displayed in Fig. \ref{fig:overtones} should be taken with a grain of salt.

\clearpage

\printglossary[type=\acronymtype]
\bibliographystyle{unsrt2}
\bibliography{ref}

\end{document}